\documentclass[aps,rmp,final,amsmath,amssymb,graphicx,longbibliography,a4paper,11pt]{revtex4-1}
\pdfoutput=1
\usepackage[version=4]{mhchem}
\usepackage{graphicx}
\usepackage[detect-all,separate-uncertainty]{siunitx}
\DeclareSIUnit\ppm{ppm}
\DeclareSIUnit\day{d}
\DeclareSIUnit\year{yr}
\DeclareSIUnit\ton{t}
\DeclareSIUnit\u{u}
\usepackage{bm}
\usepackage[english]{isodate}
\cleanlookdateon
\usepackage{subfig}
\usepackage{hyperref}

\begin{document}

\title{Detection and effects of meteoric smoke particles in the atmosphere}
\date{\printdate{25.11.2015}}
\author{Tim Dunker}
\email{tdu {at} justervesenet {dot} no}
\affiliation{Department of Physics and Technology, University of Troms\o\ -- The Arctic University of Norway, Postboks 6050 Langnes, 9037 Troms\o, Norway}
\altaffiliation{now at: National laboratory, Justervesenet, Postboks 170, 2027 Kjeller, Norway}

\maketitle

\tableofcontents

\section{Introduction}

\subsection{Scope}
This manuscript is mainly about the measurement of meteoric smoke particles in only the mesopause region by sounding rockets and, to a lesser extent, by remote sensing. Furthermore, I describe briefly some effects---either observed or simulated or both---of meteoric smoke particles on different parts of the atmosphere. I do not attempt to cover numerical simulations or results from climate--chemistry models in this manuscript. However, I refer to some of these studies at some places in the manuscript, especially in Sect. \ref{sec:properties} and Sect. \ref{sec:effects}. The references may serve as a guide for further reading.

Before looking at how we can measure meteoric smoke particles and what effects can be attributed to these particles, let us look briefly at some important terms.

\subsection{Nomenclature}
\label{sec:nomenclature}
\begin{figure}[!b]
\centering
\includegraphics[width=0.7\textwidth]{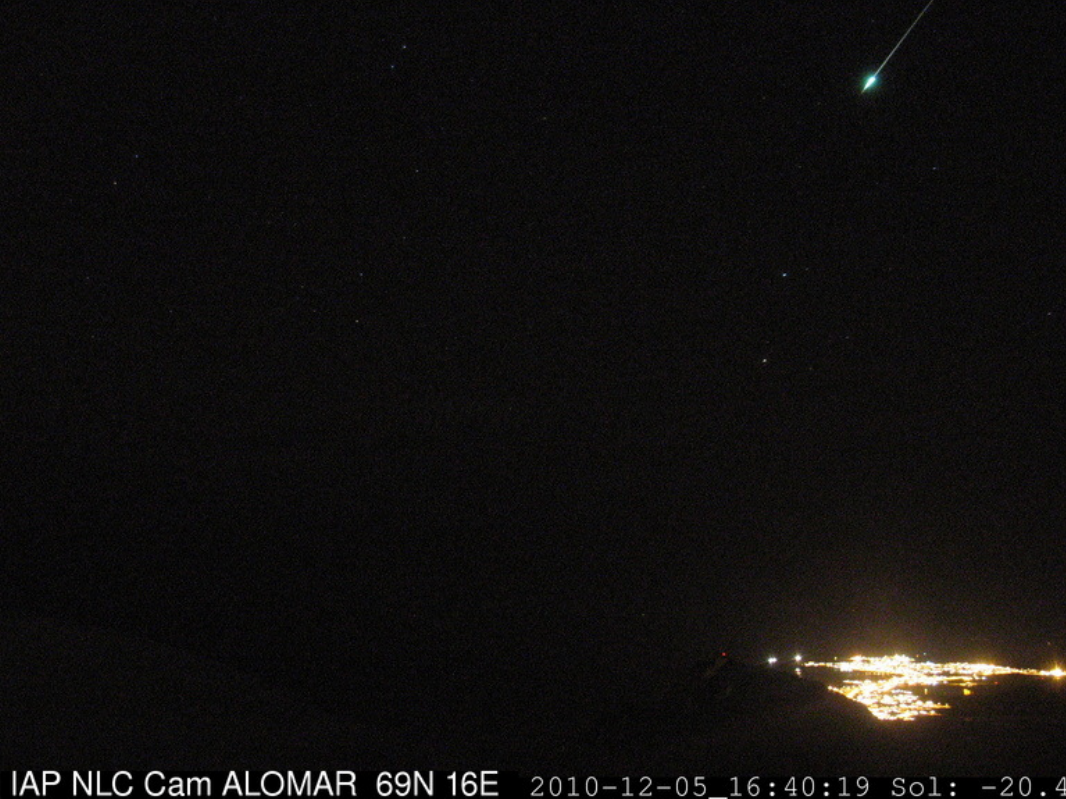}
\caption{Photograph of a meteor above Andenes ($\sim$\ang{69}N, \ang{16}E), taken by the automatic camera installed at the ALOMAR observatory, on \printdate{05.12.2010} at 16:40 UTC. The image was kindly provided by the Leibniz--Institut f\"ur Atmosph\"arenphysik an der Universit\"at Rostock e.V., who owns and operates the camera.}
\label{fig:meteor}
\end{figure}

Much of the nomenclature I use here stems from \citet{Huntenetal1980}. We need to distinguish between three terms\footnote{For a discussion on nomenclature, see \citet{RubinGrossman2010}, who proposed updated definitions of the terms ``meteoroid'' and ``meteorite'' and related terms.}, namely
\begin{itemize}
   \item a meteoroid, which is the interplanetary body that enters a planet's atmosphere and is subject to frictional heating,
   \item a meteorite, which is the solid remains of a meteoroid found on a planet's surface, and
   \item a meteor, which is the optical and/or radio phenomenon associated with a meteoroid's interaction with a planet's atmospheric constituents, see Fig. \ref{fig:meteor}.
\end{itemize}

I will not go further into the latter two. I am concerned with meteoroids only and their interaction with Earth's atmosphere. Meteoroids can be divided further into three classes:
\begin{enumerate}
   \item micrometeoroids
   \item sub--micrometeoroids
   \item superbolides
\end{enumerate}

Micrometeoroids are by far the most common bodies entering Earth's atmosphere. They are called micrometeoroids because their mass is typically in the microgram range, with their mass distribution peaking around \SI{10}{\micro\gram}.

The second class, which I call for sub--micrometeoroids, consists of bodies so small that they never sublimate or melt. They will sediment largely unaffected to lower altitudes, and eventually down to the surface. In turn, their effect on the middle atmosphere seems to be negligible.

The superbolides, the third class, are very few in number, but are typically quite spectactular. For instance, the Chelyabinsk superbolide in 2013 was so large that it created a huge fireball at tropospheric altitudes. Superbolides do not affect the middle atmosphere much, though.

\begin{figure*}[!tb]
\centering
\includegraphics[width=0.7\textwidth]{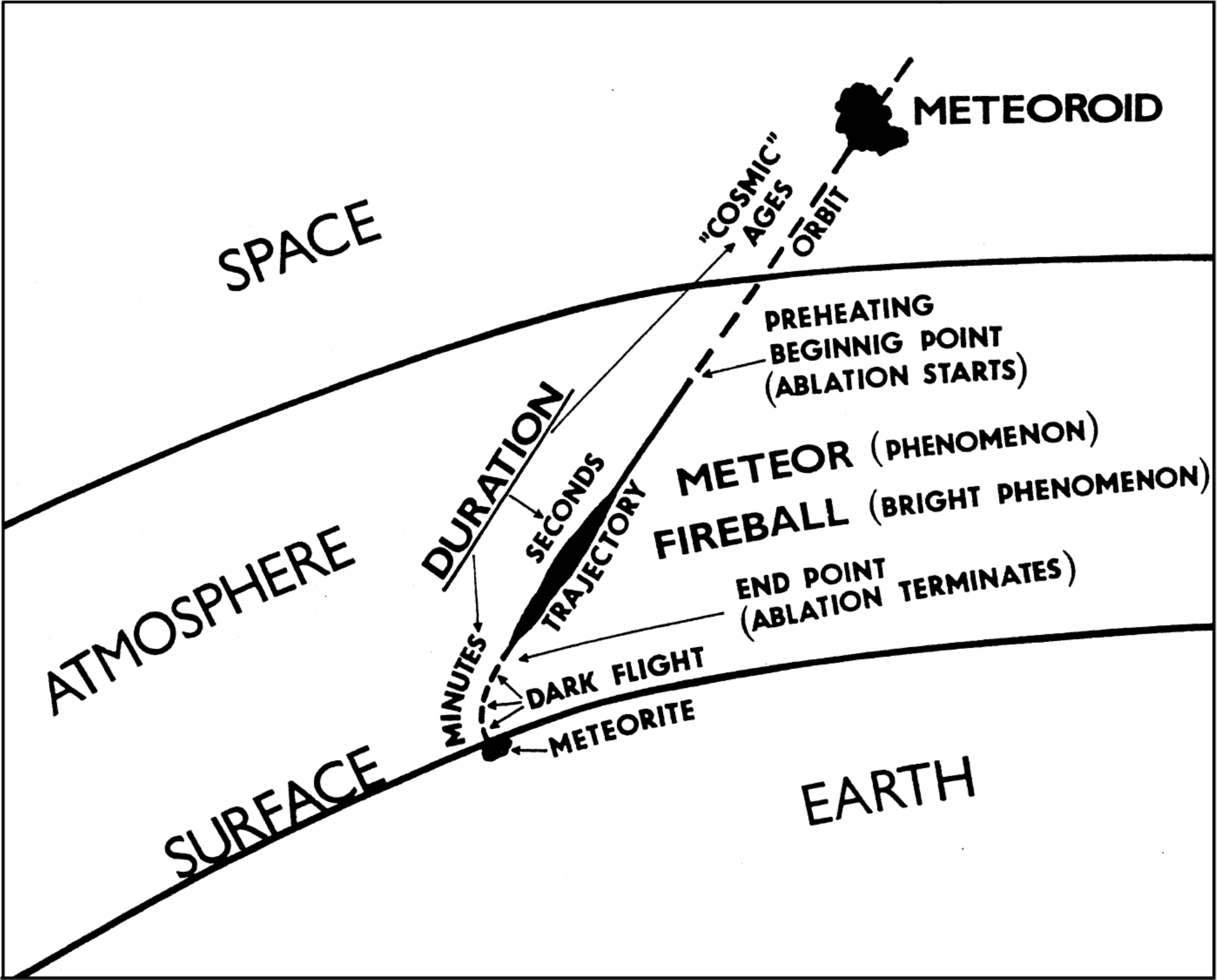}
\caption{Basic processes and nomenclature of meteoroid interaction with Earth's atmosphere. Figure from \citet[][Fig. 2]{Ceplechaetal1998}.}
\label{fig:ceplechaetal1998}
\end{figure*}

Incoming particles interact with Earth's atmosphere. Figure \ref{fig:ceplechaetal1998} is a schematic overview of the processes that an incoming meteoroid may undergo. Meteoroids enter Earth's atmosphere with velocities between \SIrange{11}{72}{\kilo\metre\per\second}, depending on the orientation of motion (prograde or retrograde). Any incoming meteoroid experiences frictional heating through its collision with air molecules. This frictional heating supplies the energy needed for sublimation. The whole process is called meteoric ablation. In order to understand what the resulting atmospheric effects might be, let us look closer at how much material we speak of and what it is composed of.

The seemingly easy question of how much material meteoroids in these three classes ablate in the atmosphere has proven difficult to answer. Estimates range from \SIrange[range-phrase=\ to\ ]{2}{300}{\ton\per\day}, and different observational methods yield different estimates. The mass influx is distributed over \num{30} orders of magnitude: from \SIrange[range-phrase=\ to\ ]{1e-18}{1e12}{\kilogram}. The largest masses are the superbolides, which are very rare. Most meteoroids have masses between \SIrange[range-phrase=\ and\ ]{1e-12}{1e-6}{\kilogram}, with a mode value of \SI{1e-8}{\kilogram}. Meteoroids that cause meteors (i.e. the visible phenomenon) have typical masses of $\sim$\SI{1e-5}{\kilogram}. Meteoroids consist of several elements (e.g. \ce{Fe}, \ce{Mg}, \ce{Na}, \ce{Si}, \ce{K}, \ce{Ca}), which are subject to differential ablation, that is, they ablate at different altitudes, but generally between \SIrange[range-phrase=\ and\ ]{70}{110}{\kilo\metre}. In fact, it is the sublimation temperature of an element that causes some elements to ablate differentially. Section \ref{sec:composition} deals with the composition of meteoric smoke particles---which is different from that of meteoroids.

\subsubsection{Are meteoric debris and meteoric smoke particles the same thing?}
Originally, the title of the lecture on which this manuscript is based, included the term ``meteoric debris'' instead of ``meteoric smoke particles''. The terms, as I regard them, are not synonymous. So, the short answer to the question posed in the title is ``no''. I prefer the latter term for several reasons.

The term ``meteoric debris'' is not a common one; it appears to me that it has been used only a couple of times over the last three decades or so. When authors wrote ``meteoric debris'', they always refrained from defining what they meant by it. Instead, meteoric debris seems to have been used as a synonym for either ``secondary meteoric material'' or ``meteoric smoke particles'', for example, simply to avoid repetition. However, the three terms do not mean the same thing, which warrants clarification and brings about the question of what is considered primary meteoric material.

I argue that primary meteoric material is the material that enters Earth atmosphere and which may be ``split'' into its constituents, without either interacting chemically with atmospheric constituents or forming new particles through coagulation. Then, secondary meteoric material is particles that are of meteoric origin---at least in part---and which have undergone chemical interaction with atmospheric constituents or have coagulated to form larger particles. \citet{RosinskiSnow1961} proposed a theory of the formation of secondary meteoric particles. \citet{Huntenetal1980} elaborated on this theory and modelled the processes. Still, it took \num{15} years before such particles were detected in the atmosphere \citep{Havnesetal1996}. Following the nomenclature of \citet{Huntenetal1980}, I call these secondary, solid particles ``meteoric smoke particles''. These meteoric smoke particles\footnote{Sometimes, meteoric smoke particles are called ``cosmic dust''. A ``dust particle'' is an inclusive term for any solid particle entering the atmosphere or already in it.}, are arguably very important for the mesospheric chemistry and ionospheric charge balance.

A rough search on the Web of Knowledge\footnote{\href{apps.webofknowledge.com}{apps.webofknowledge.com}} revealed that, since 1981, the term ``meteoric debris'' has been used three times as part of a manuscript title and eleven times as a topic designation. The term ``meteoric smoke'' has appeared 32 times and 101 times, respectively. The term ``cosmic dust'' appeared in article titles 368 times, but the vast majority of the appeareances are within astronomy, astrophysics, and geophysics. About 50 appearances were in the field of atmospheric physics. However, the term ``meteoric smoke'' better reflects the processes that lead to the formation of the particles. Cosmic dust and meteoric smoke are not a synonym, it seems to me.

Generally, meteoric debris can be regarded as a much broader category, including all solid material of meteoric origin. Meteoric smoke particles is a much narrower category, including a clearly defined process through which these particles come into being. Of all meteoric debris, meteoric smoke particles seem to have the most important impact on Earth's atmosphere.

\subsection{Mesosphere and D region of the ionosphere}
\begin{figure*}[!t]
  \centering
  \includegraphics[width=0.7\textwidth]{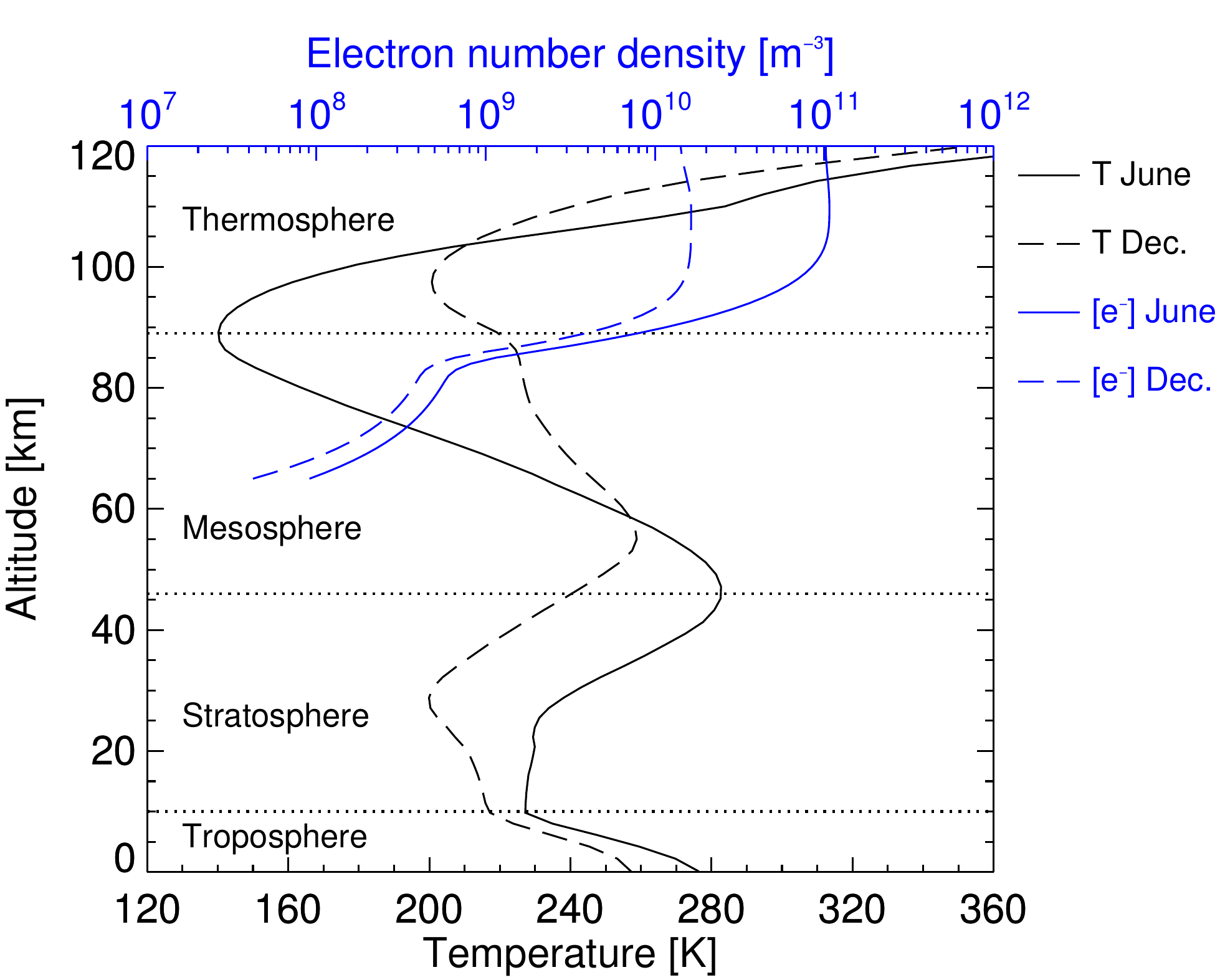}
  \caption{Vertical profiles of mean June (solid black line) and December (dashed black line) temperature from the COSPAR International Reference Atmosphere 1986. The mean June (solid blue line) and December (dashed blue line) electron number density are taken from the International Reference Ionosphere.}
  \label{fig:tempelectrons}
\end{figure*}
To explain the observations, knowledge of the composition of the background atmosphere is essential. The mesosphere is the layer of the neutral atmosphere that extends from the stratopause ($\sim\SI{50}{\kilo\metre}$) to the mesopause ($\sim$\SIrange[range-phrase=\ to\ ]{80}{110}{\kilo\metre}). The summer polar mesopause is the coldest part of Earth's atmosphere, and is host to noctilucent clouds, which are the highests clouds in the atmosphere. The mesopause region is also the region where meteoroids interact with the atmospheric constituents, making possible the formation of meteoric smoke particles. 

Figure \ref{fig:tempelectrons} shows vertical profiles of mean neutral temperature for June and December at \SI{70}{\degree}N up to an altitude of \SI{120}{\kilo\metre}, and typical electron number density profiles for these months. Note that there is a coexistence of neutral and charged particles in the mesosphere and lower ionosphere. This will be important to explain the occurrence of charged particles.

\section{Formation and properties of meteoric smoke particles}
\label{sec:properties}

\subsection{Formation and growth}
\label{sec:formation}
Very much of our knowledge about the formation and the properties meteoric smoke particles relies on model and laboratory results. According to \citet[][p. 1]{Huntenetal1980}, ``smoke particles are immediate condensation products within a meteor trail and their immediate descendants through coagulation and sedimentation''.  Through collisions with air molecules, all or parts of a meteoroid sublimates. If a meteoroid is very small (a sub--micrometeoroid, see Sect. \ref{sec:nomenclature}), it might not undergo sublimation at all, because the temperature never reaches the melting point. From their model study of meteoric smoke production, \citet{Kalashnikovaetal2000} found that about \SI{75}{\percent} of all meteoroids ablated entirely. A meteoroid's sublimated constituents can recombine whenever they collide. More specifically, meteoric smoke forms through the polymerization of metal hydroxides, hydroxides, carbonates, and silicon dioxide particles (J. M. C. Plane, pers. comm., 2013). The recombination of meteoroid ablation products does not lead to a heat excess large enough to cause immediate dissociation. In addition, the kinetic energy of molecules at ambient upper mesospheric temperatures is too low to supply the necessary energy for dissociation. The vapour pressure of the constituents is so low that there is no need for condensation nuclei---all molecules act as a condensation nuclei \citep{RosinskiSnow1961}. That is, once formed, the meteoric smoke particles are not easily removed from the mesopause region. The particles need about ten days to grow to a diameter of $\sim$\SI{1}{\nano\metre} (J. M. C. Plane, pers. comm., 2013).

\begin{figure*}[!bt]
  \centering
  \includegraphics[width=0.7\textwidth]{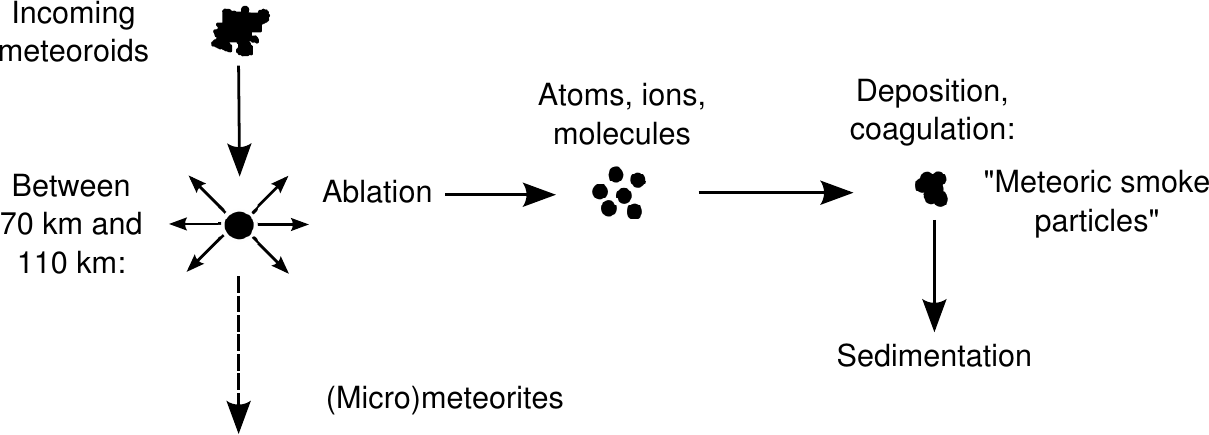}
  \caption{Schematic overview of the processes that lead to the formation of meteoric smoke particles. Adapted from the original figure made by J\"org Gumbel, Meteorologiska institutionen, Stockholms universitet, Sweden.}
  \label{fig:mspformation}
\end{figure*}

\subsection{Composition and size}
\label{sec:composition}
According to \citet{GomezMartinPlane2013}, there are two types of meteoric smoke particles; those that contain silicon (metal silicate particles), and those that do not contain silicon (metal hydroxides). For particles with $r<\SI{1}{\nano\metre}$, \SI{66}{\percent} are metal hydroxides and, thus, they do not contain silicon.

The results of \citet{GomezMartinPlane2013} are in agreement with results of \citet{Rappetal2012}, who ruled out silicon as a constituent of their detected meteoric smoke particles, and argued that metal hydroxides (\ce{MgOH} and \ce{FeOH}) are likely candidates. This judgement was based on measurements of the work function of meteoric smoke particles.

However, silicon may still be an important atmospheric constituent in the formation of meteoric smoke particles. \citet{Planeetal2016} modelled silicon chemistry, and found that a likely silicon sink is \ce{Si(OH)4}, which can react with metal hydroxides to form meteoric smoke particles. 

\begin{figure*}[!t]
   \centering
   \subfloat[][]{\label{fig:msp_candidates}\includegraphics[width=0.43\textwidth]{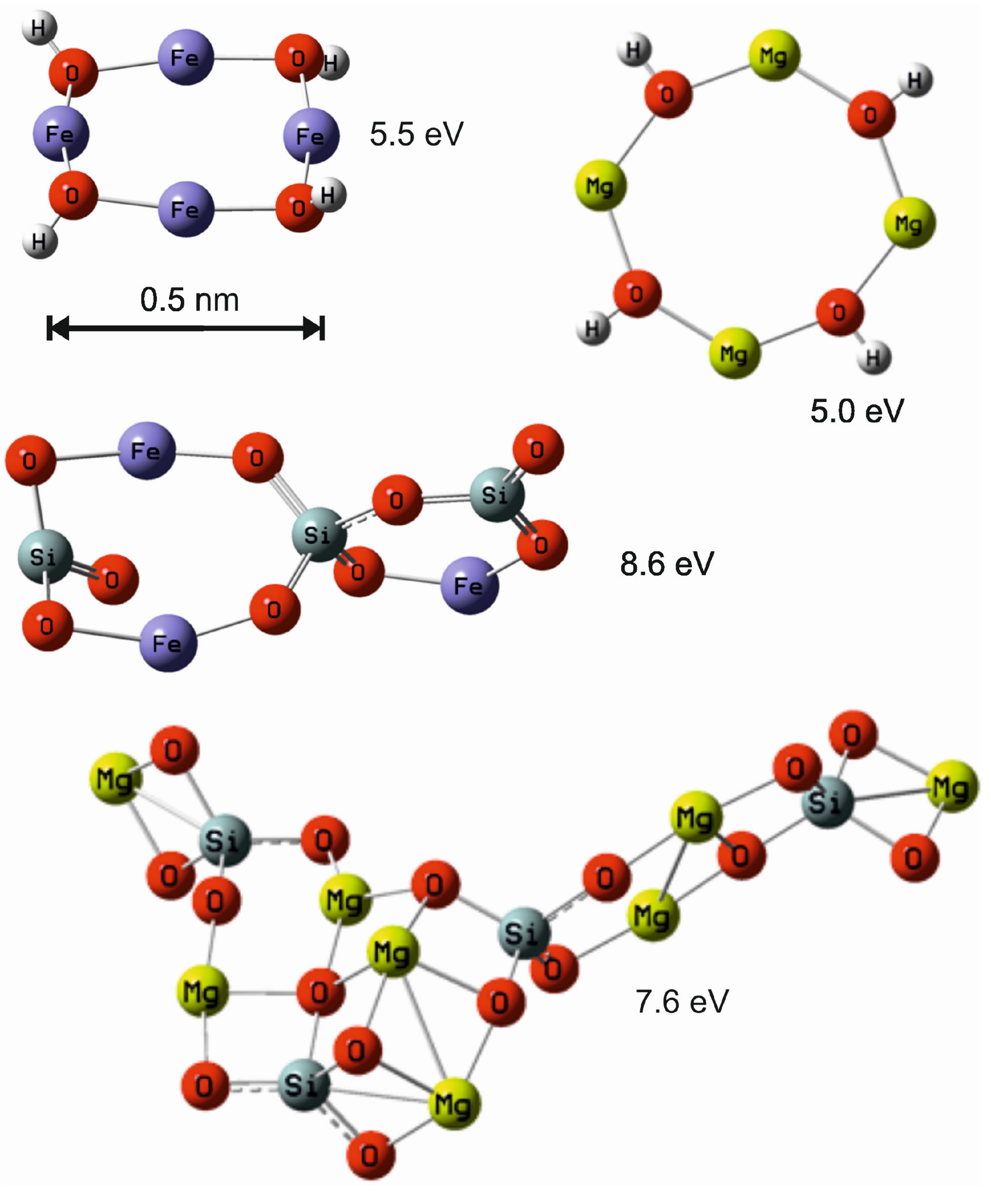}}
   \subfloat[][]{\label{fig:msp_workfunction}\includegraphics[width=0.53\textwidth]{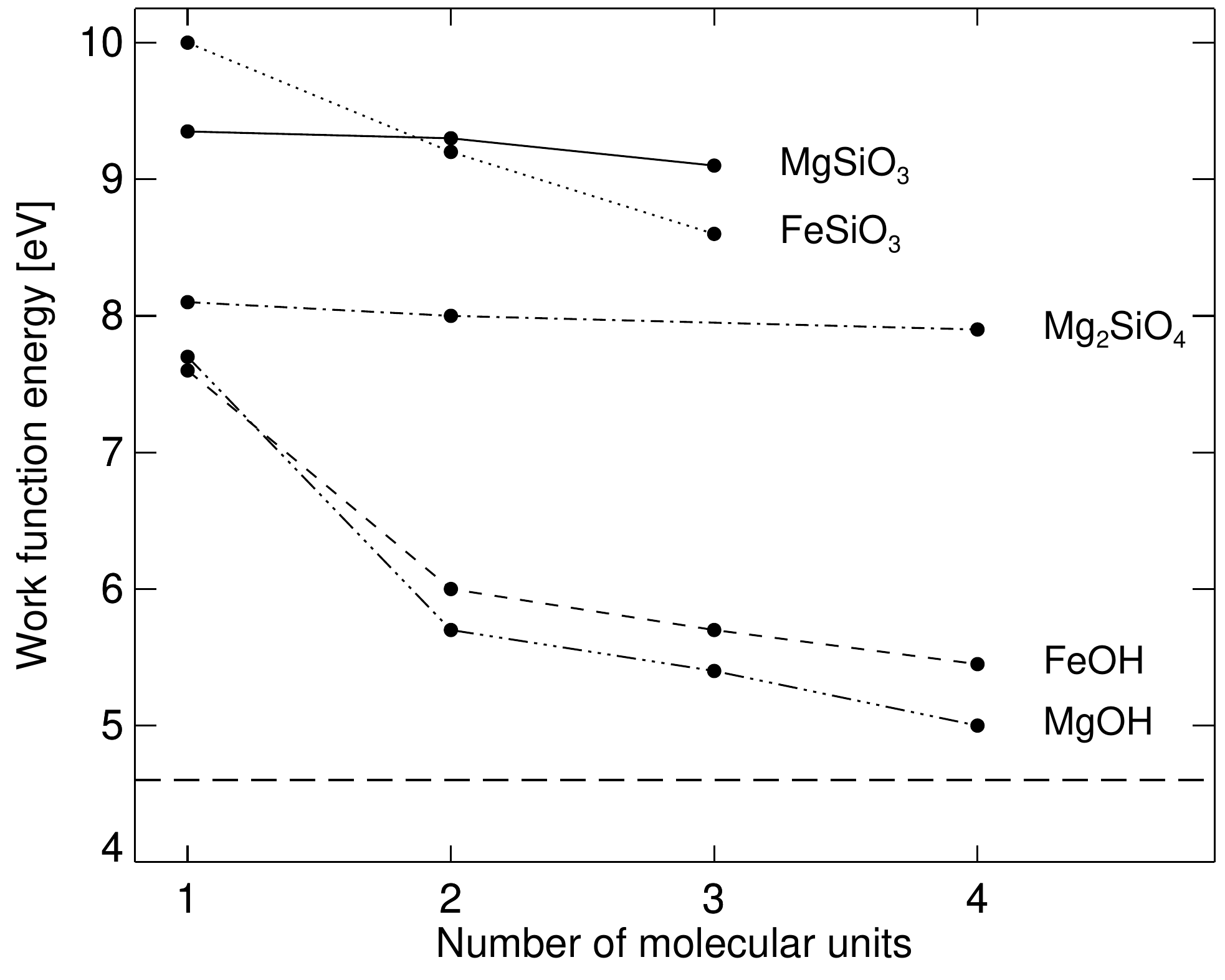}}
   \caption{(a) Chemical structure and composition of proposed meteoric smoke particle candidates, along with the corresponding work function. The molecules are \ce{(MgOH)4}, \ce{(FeOH)4}, \ce{(FeSiO3)3}, and \ce{(Mg2SiO4)4}. The ``diameter'' of \ce{(FeOH)3} is aproximately \SI{0.5}{\nano\metre}. Figure reprinted from \citet[][Fig. 10]{Rappetal2012}. (b) Work function energy as a function of cluster size for the particles shown in panel (a). Figure reprinted from \citet[][Fig. 11]{Rappetal2012}.}
\end{figure*}

Figure \ref{fig:msp_candidates} shows the structure of some proposed meteoric smoke particle candidates. The corresponding work function is also shown. The work function of a given particle depends on the cluster size, as shown in Fig. \ref{fig:msp_workfunction}: the larger the particle (the more molecular units), the lower the work function.

\begin{table}[!t]
\centering
\caption{Atomic mass for some likely meteoric smoke particle compositions. $\SI{1}{u} \approx \SI{1.66e-27}{\kilogram}$}.
\label{tab:atomicmass}
\begin{ruledtabular}
\begin{tabular}{lll}
Chemical composition & Atomic mass / \si{u} & Atomic mass / \si{\kilogram} \\
\cline{1-3}
\ce{MgSiO3} & \num{100} & \num{166e-27}\\
\ce{(MgOH)4} & \num{164} & \num{272e-27}\\
\ce{(FeOH)4} & \num{292} & \num{485e-27}\\
\ce{(FeSiO3)3} & \num{396} & \num{657e-27}\\
\ce{(Mg2SiO4)4} & \num{560} &  \num{930e-27}\\ 
\end{tabular}
\end{ruledtabular}
\end{table}

Atomic masses of some of the proposed meteoric smoke particles are listed in Table \ref{tab:atomicmass}. The mass of particles are important for the interpretation of experimental results, see Sect. \ref{sec:massspec}. One often needs the mass density of the particle rather than the atomic mass. Unfortunately, the mass density is poorly known, if at all, and it is probably not constant. Usually, a mass density of $\varrho = \SI{2e3}{\kilogram\per\cubic\meter}$ of meteoric smoke particles is assumed in experimental and model studies, for example by \citet{SaundersPlane2006}, \citet{SaundersPlane2011}, \citet{Planeetal2014}, and \citet{Robertsonetal2014}.

\begin{figure*}[!t]
  \centering
  \includegraphics[width=0.7\textwidth]{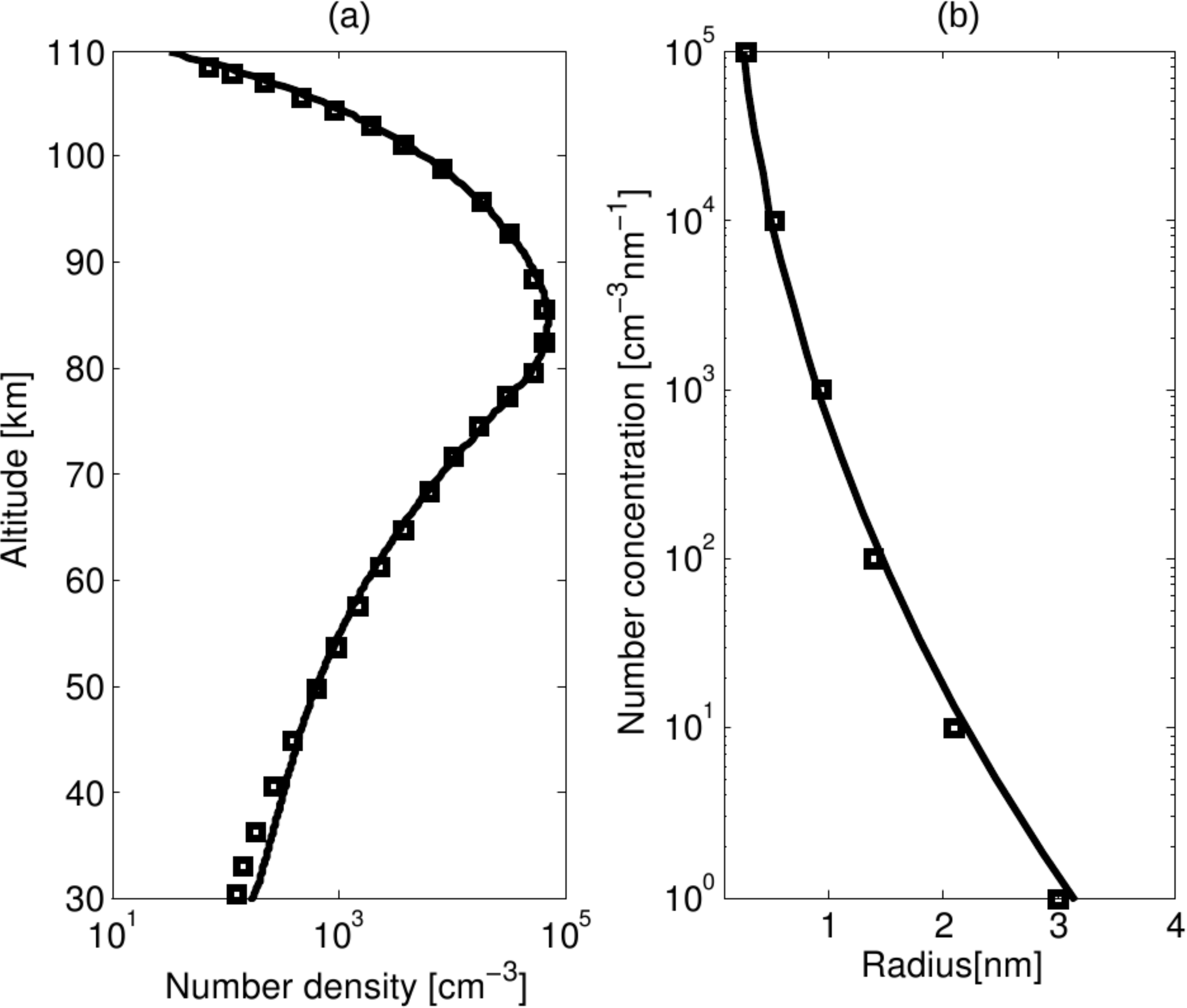}
  \caption{Simulation results of meteoric smoke properties. Refer to the text for the underlying assumptions. (a) Vertical profile of meteoric smoke particle number density, given in $\si{\per\cubic\centi\metre} = \SI{1e-6}{\per\cubic\metre}$. (b) Meteoric smoke particle size distribution for several effective radii in \si{\nano\metre}. For comparison, the squares are the values of the original simulation by \citet{Huntenetal1980}. Figure reprinted from \citet[][Fig. 2]{Megneretal2006}.}
  \label{fig:megneretal2006fig2}
\end{figure*}

A vertical profile of simulated meteoric smoke particle number density and a semilogarithmic meteoric smoke particle size distribution from \citet{Megneretal2006} is shown in Fig. \ref{fig:megneretal2006fig2}. \citet{Megneretal2006} performed a simulation similar to the one by \citet{Huntenetal1980}, with the same input parameters: an initial meteoric smoke particle radius of $r_0=\SI{0.2}{\nano\metre}$ (roughly a silicon molecule), no vertical wind, coagulation by Brownian motion, and the same meteoric ablation profile. The important panel at this point is panel (b): note how the number concentration falls off rapidly with increasing particle radius. The smallest particles, with radii smaller than \SI{1}{\nano\metre} make up the vast majority of particles in the simulation.

\subsection{Charge state}
Recall from Fig. \ref{fig:tempelectrons} that free electrons exist between \SIrange[range-phrase=\ and\ ]{70}{110}{\kilo\metre}, where much of the ablation takes place. Therefore, I might expect that some meteoric smoke particles might carry a charge. In fact, this is what is observed. Both negatively-- and positively--charged meteoric smoke particles have been observed. 

The dominant charging process for negatively charged meteoric smoke particles appears to be electron attachment. In the D region of the ionosphere, free electrons have a thermal velocity much larger than that of heavier ions or meteoric smoke particles. The charging rate depends on the particle's size: the larger the particles, the higher is the rate of charging \citep{Rapp2000}. Therefore, large meteoric smoke particles should be charged negatively.

\citet{Rapp2009} showed that for small meteoric smoke particles, the rate of photodetachment and photoionization is much higher than the electron--capture rate, unless the electron number density is very large. Therefore, under sunlit conditions, small meteoric smoke particles should be positively charged or neutral. His calculations explain the coexistence of positive and negative meteoric smoke particles in noctilucent clouds, because the smallest meteoric smoke particles can only be positively charged or neutral \citep{Rapp2009}. The larger meteoric smoke particles and ice particles must be negatively charged, because photodetachment and photoionization are negligible for these \citep{Rapp2009}.

\subsection{Abundance and transport}
\label{sec:abundance}

\begin{figure}[!t]
 \centering
 \includegraphics[width=0.7\textwidth]{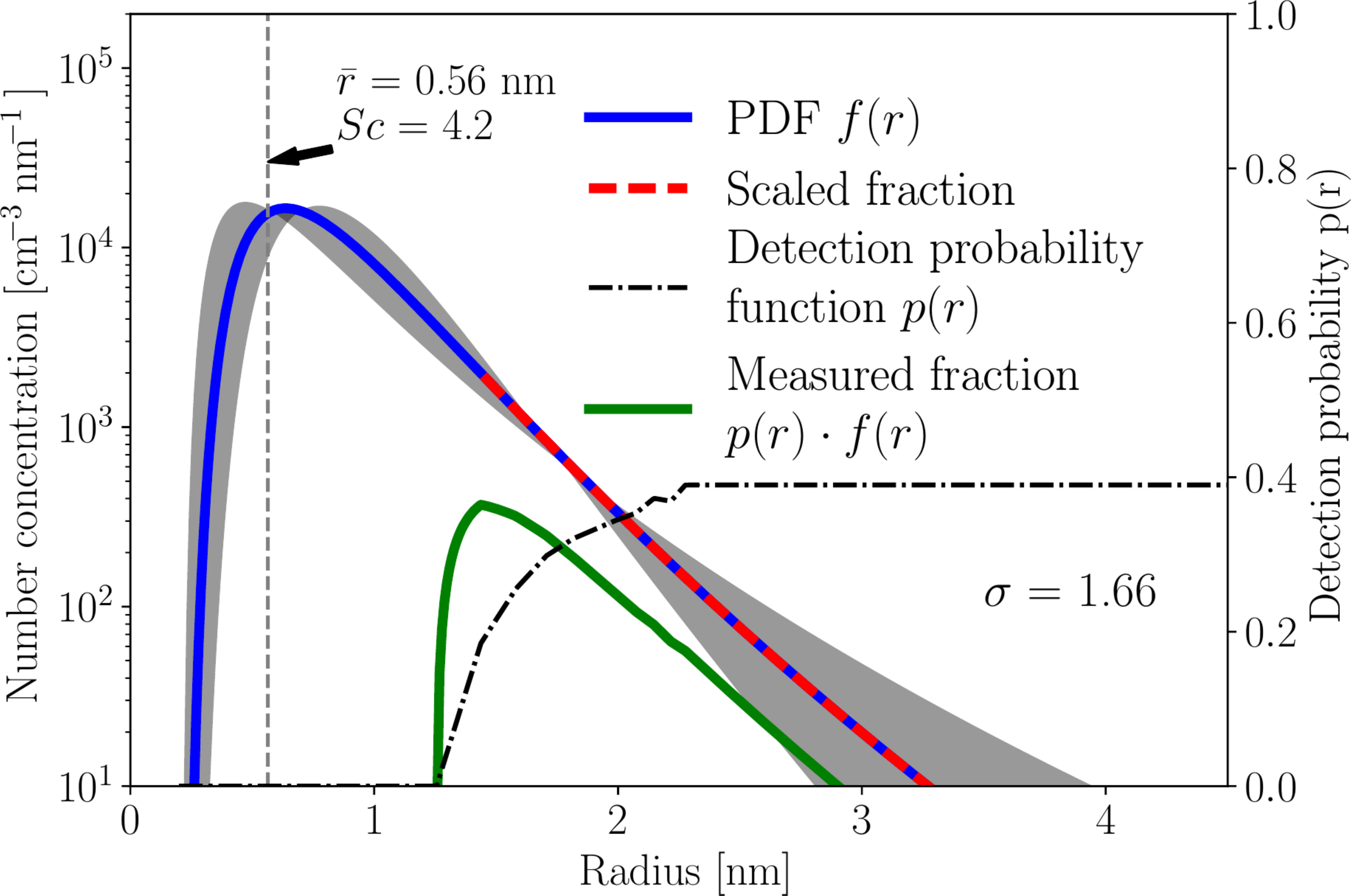}
 \caption{Number concentration and detection probability as a function of particle radius. The width of the measured fraction (green) is given by $\sigma = \SI{1.66}{\nano\metre}$. The mean radius of meteoric smoke particles at \SI{82}{\kilo\metre} was $\bar{r} = \SI{0.56}{\nano\metre}$, derived from Schmidt number measurements. These are experimental results. Figure adapted from \citet[][Fig. 15]{Asmusetal2017}. See \citet{Asmusetal2017} for details.}
 \label{fig:asmusetalfig15}
\end{figure}

Figure \ref{fig:asmusetalfig15} shows the size--dependent distribution function of meteoric smoke particles at \SI{82}{\kilo\metre}, measured by the WADIS--2 sounding rocket, which was launched from And\o ya (\ang{69}N) in March 2015 \citep{Asmusetal2017}. The figure requires some explanation. As is customary in cloud physics, also here the size--distribution function is described by a log--normal function. The green, solid line is the fraction of meteoric smoke particles actually measured by the particle detector. This function is the convolution of the log--normal distribution (blue line) and the detection probability (black dash--dotted line). That is, no particles smaller than about \SI{1.2}{\nano\metre} were detected. Note the left y--axis scale: it is not such that no particles larger than about \SI{3}{\nano\metre} were detected. Because the detection probability is non--zero, the number density of such particles was smaller than \SI{1e-1}{\per\cubic\centi\metre\per\nano\metre}.
The figure shows that small particles are much more abundant than large particles, with a mean radius of \SI{0.56}{\nano\metre} and a corresponding number density of roughly \SI{1e4}{\per\cubic\centi\metre} at \SI{82}{\kilo\metre}. The mean radius for the region between \SIrange[range-phrase=\ and\ ]{81}{85}{\kilo\metre} was slightly smaller: $r=\SI{0.48}{\nano\metre}$ \citep{Asmusetal2017}. Note that this is not a direct measurement of meteoric smoke particle properties. The technique is based on spectral analysis of electron number density profile, from which the particle size is derived via Schmidt number profiles\footnote{For more details, see \citet{Luebkenetal1994}, \citet{Luebkenetal1998}, and \citet{Asmusetal2017}.}.

First of all, meteoric smoke particles are not a major atmospheric constituent. Panel (a) of Fig. \ref{fig:megneretal2006fig2} shows one vertical profile, extending from the middle stratosphere up to the lower thermosphere, of meteoric smoke particle number density. Around the mesopause, meteoric smoke particles are most abundant with number densities just below \SI{1e11}{\per\cubic\metre} (a rather large figure in light of some experimental results). Above and below the mesopause, the number density decreases down to about \SI{1e3}{\per\cubic\centi\metre} at \SI{30}{\kilo\metre} and \SI{107}{\kilo\metre}. Typical air number densities at mesopause altitudes are $\sim$\SI{1e20}{\per\cubic\metre}. The number density of meteoric smoke particles is comparable to the number density of the plasma in the D region, typically $\sim$\SI{1e9}{\per\cubic\metre}. That is, meteoric smoke particles are a trace species\footnote{They are only trace species above the stratopause. In the stratosphere, the particles can no longer be considered passive tracers \citep{Hervigetal2017}.}, but still non--negligible constituents, see Sect. \ref{sec:effects}. 

The single profile from Fig. \ref{fig:megneretal2006fig2} is very useful to give us some understanding of the abundance in space, but not in time, of course---we cannot assume the number density of meteoric smoke particles to be constant. For example, the polar summer mesosphere contains many fewer meteoric smoke particles than the polar winter atmosphere does. The reason is that the meteoric smoke particles are transported to the winter pole by the residual meridional circulation of the mesosphere.

\begin{figure*}[!b]
   \centering
   \includegraphics[width=0.7\textwidth]{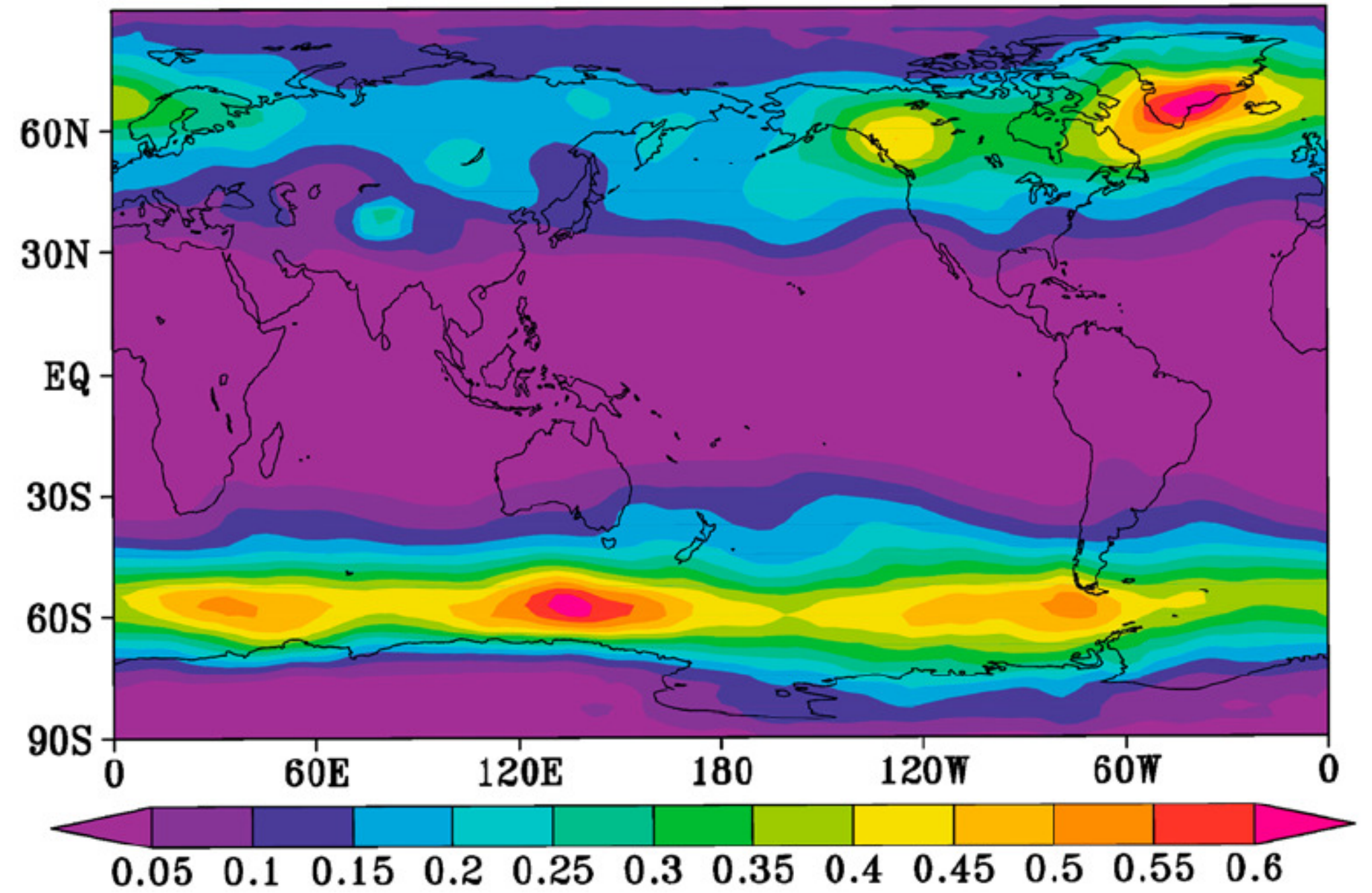}
   \caption{Latitude--longitude plot showing the annual mean deposition of iron, in \SI{1e-6}{\mol\per\square\metre\per\year}. Figure reprinted from \citet[][Fig. 3(b)]{Dhomseetal2013}.}
   \label{fig:dhomseetal2013}
\end{figure*}

As remarked in Sect. \ref{sec:formation}, meteoric smoke particles form through deposition and coagulation in the mesopause region. Initially, these particles are of sub--nanometre size, but can grow to nanometre size in the course of about ten days, as mentioned earlier (J. M. C. Plane, pers. comm., 2013). Meteoric smoke particles with radii $r_{eff} < \SI{5}{\nano\metre}$ do not sediment quickly below about \SI{80}{\kilo\metre}, but are transported with the residual meridional circulation of the mesosphere to the winter pole \citep{Plane2012}. The latitudinal distribution of meteoric smoke particles therefore exhibits a maximum concentration in the winter polar mesosphere and stratosphere, and a minimum in the corresponding summer layers. At tropical latitudes, the annual variation is small.

Vertical transport---more precisely, meridional transport to high latitudes where downward motion takes place---to tropospheric altitudes raises the question of the fate of meteoric smoke particles. \citet{Dhomseetal2013} modelled the annual mean deposition of iron, see Fig. \ref{fig:dhomseetal2013}. They found that there are relative maxima at the Southern Ocean surface and over southern Greenland \citep{Dhomseetal2013}. In general, \citet{Dhomseetal2013} found the deposition pattern to be quite uniform over the Southern Ocean compared with the Arctic. It is surprising to see that the deposition does not seem to take place over the poles. The residual meridional circulation of the mesosphere transports particles to the poles, where the particles sediment. Therefore, one would expect the deposition to be larger there than in the latitude band between \ang{30} and \ang{70}, cf. Fig. \ref{fig:dhomseetal2013}. The reason for this deposition distribution is not yet understood.

\section{Effects of meteoric smoke particles on atmospheric regions}
\label{sec:effects}
Despite their relatively small concentration (regardless of season or latitude), meteoric smoke particles have important effects on the atmosphere. The nature of the effect depends on altitude, and on the particles' charge state. What follows is an overview, from high altitudes to low altitudes, of effects as we know them today.

\subsection{Mesosphere}

\subsubsection{Charge balance}
\begin{figure*}[!t]
  \centering
  \includegraphics[width=0.7\textwidth]{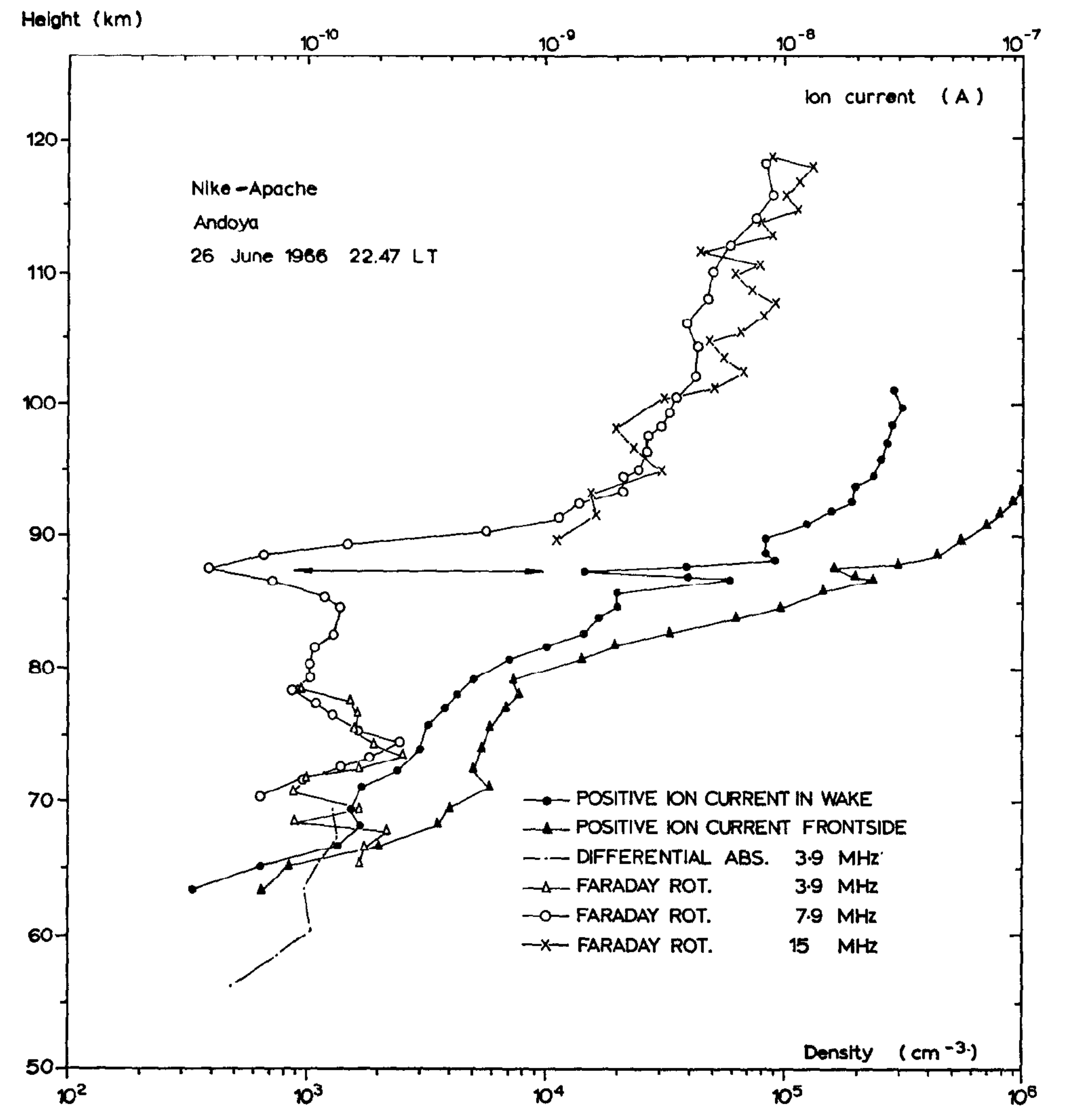}
  \caption{Sounding rocket measurements of positive ions and electrons. The measurements were made on And\o ya on \printdate{26.06.1966}, that is, in polar summer. The number density is given in \si{\per\cubic\centi\metre}, the measured current is in \si{\ampere}. Figure reprinted from \citet[][Fig. 1]{Pedersenetal1970}.}
  \label{fig:pedersenetal1970}
\end{figure*}

Starting in the mesosphere/lower ionosphere region, two effects have been debated and investigated. As we have already seen, meteoric smoke particles can be charged. Therefore, their presence can alter the charge balance of the D region of the ionosphere. For example, electron bite outs---layers in which electrons are much less abundant than just above and below this layer---have been observed in sounding rocket experiments \citep[for instance,][]{Pedersenetal1970,Ulwicketal1988}. \citet{Pedersenetal1970} showed that electrons can be depleted in the presence of ice particles, see Fig. \ref{fig:pedersenetal1970}. Using their sounding rocket data, they observed a decrease in electron number density in polar summer above the noctilucent cloud altitudes ($\sim$ \SIrange[range-phrase=\ to\ ]{80}{85}{\kilo\metre}). A possible mechanism for these electron bite--outs has been identified in electron attachment to meteoric smoke particles \citep[e.g.][]{Friedrichetal2009,Friedrichetal2012}. \citet{Rappetal2005} launched a sounding rocket with instruments that measured free electrons, positive ions, and the current created by charged meteoric smoke particles. Their results showed that the number density of free electrons was not balanced by the number density of positive light ions, and they attributed the difference to the existence of positively charged meteoric smoke \citep{Rappetal2005}.

\subsubsection{Ice nuclei}
\label{sec:nuclei}
At first glance, it may seem surprising that ice particles exist at altitudes up to \SI{90}{\kilo\metre}. It is well--known that the polar summer mesopause is the coldest part of Earth's atmosphere, and under certain conditions icy clouds may form. These clouds, at altitudes between about \SIrange[range-phrase=\ and\ ]{81}{85}{\kilo\metre}, have been known since at least 1885. The ice particles can be visible to the naked eye, if they have reached effective radii of $r_{eff} \sim \SI{10}{\nano\metre}$ and are then termed noctilucent clouds\footnote{The term ``noctilucent cloud'' is most often used in studies that use ground-based observations. The use of the term does not mean to imply that the clouds are luminous themselves. They are illuminated by the sun from below.} or polar mesospheric clouds.

\begin{figure*}[!b]
  \centering
  \includegraphics[width=0.7\textwidth]{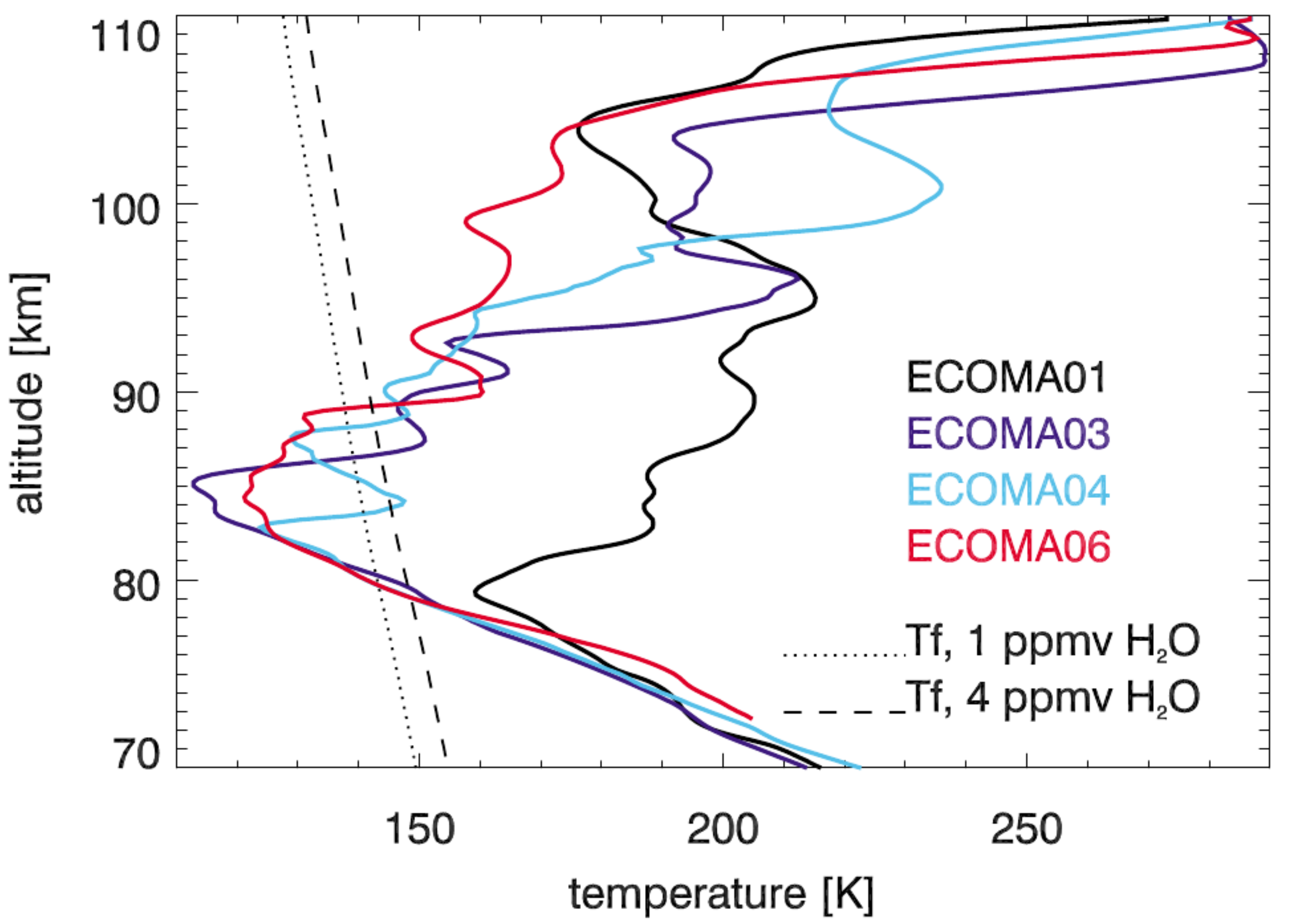}
  \caption{Temperature measurements made during four sounding rocket launches. The measurements were all made from And\o ya. Figure reprinted from \citet[][Fig. 6]{Rappetal2010}.}
  \label{fig:rappetal2010fig6}
\end{figure*}

Temperature measurements made by instruments on sounding rockets have shown that temperature do drop below the freezing point of water vapour \citep{Rappetal2010}. Figure \ref{fig:rappetal2010fig6} shows temperature profiles measured during the ECOMA campaign and the frost point of water vatour for two different volume mixing ratios. In three of these profiles, the temperature is colder than the frost point of water vapour.

Even though the temperature of the ambient atmosphere might be colder than the frost point for a given water vapour mixing ratio, the required energy for water vapour to undergo deposition (i.e., form ice crystals instantaneously) is too large to be overcome by homogeneous processes, given the observed levels of supersaturation with respect to water vapour. Ice particles may form, however, if ice nuclei are present. This leads to the question: which properties must a particle possess to become an ice nuclei?

There are basically two essential characteristics:
\begin{enumerate}
 \item larger than a critical radius
 \item relatively large dipole moment
\end{enumerate}

According to thermodynamics, a system will always tend to minimize its free energy (Gibbs free energy). For nucleation on meteoric smoke particles to occur, the meteoric smoke particles need to be larger than a critical radius\footnote{The term ``radius'' is an approximation, because ice particles are not spherical, such that a classical ``radius'' is not defined.}, above which the free energy decreases. The higher the supersaturation, the smaller the critical radius under otherwise the same conditions.

To be an effective ice nucleus, a particle must have a rather large electric dipole moment, because water molecules need to bind easily to the ice nucleus. The reason is the free--energy barrier is greatly reduced \citep[][p. 4525]{Planeetal2015}.

Certain meteoric smoke particles possess either one or both of the characteristics just described. Especially a negatively charged meteoric smoke particle may act as an ice nucleus, because it has a large dipole moment, and therefore binds easily to water molecules. Still, also some neutral meteoric smoke particles $\left(\ce{FeSiO3} and \ce{MgSiO3}\right)$ may act as ice nucleim, because these also have large dipole moment \citep{GumbelMegner2009}.

\begin{figure*}[!t]
  \centering
  \includegraphics[width=0.7\textwidth]{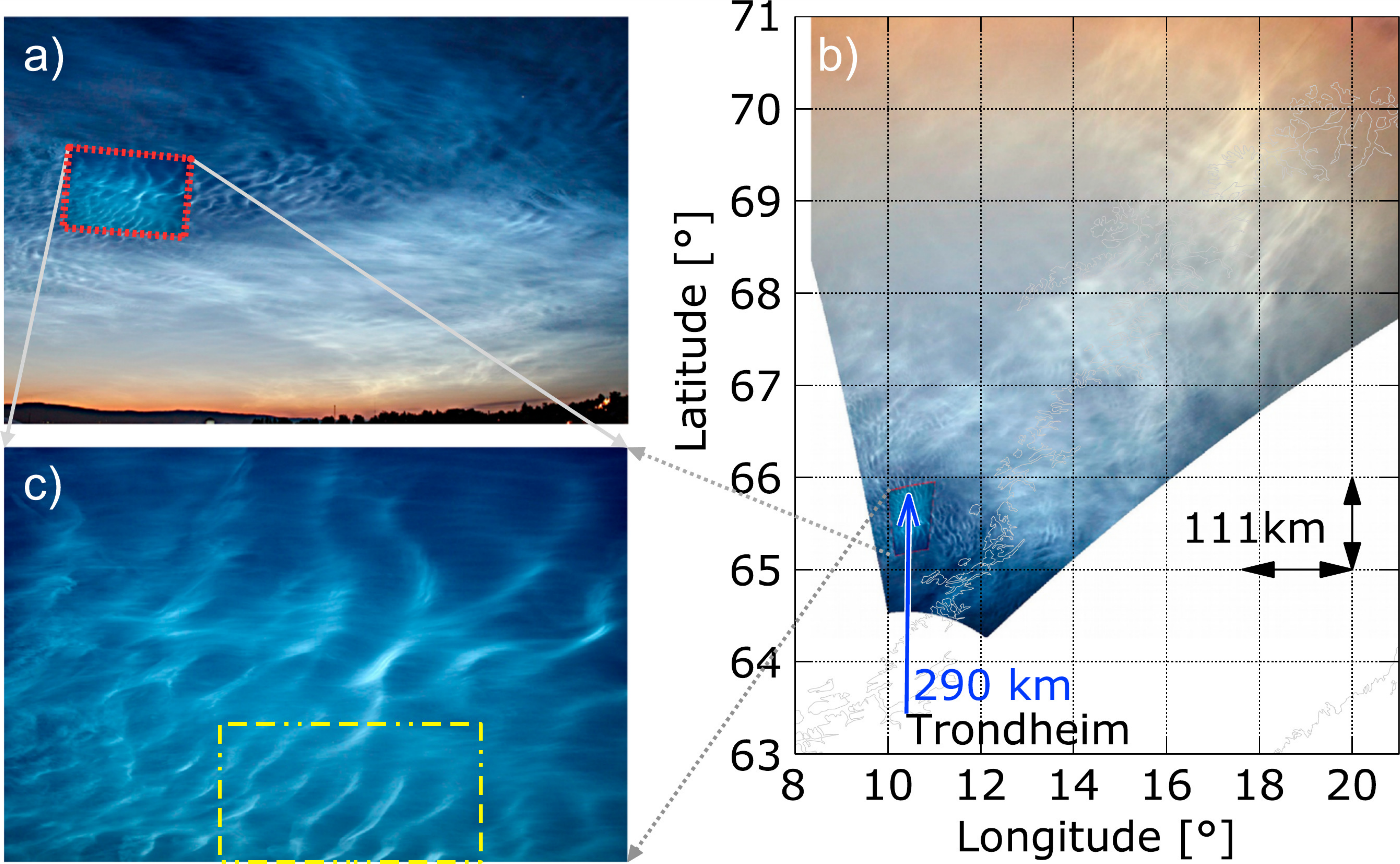}
  \caption{Photographs of polar mesospheric clouds, taken on \printdate{01.08.2009} at 23:00. The clouds are at about \SI{83}{\kilo\metre} altitude. The yellow rectangle indicates an area where possible Kelvin--Helmholtz instabilities occurred. Figure reprinted from \citet[][Fig. 7]{BaumgartenFritts2014}.}
  \label{fig:baumgartenfritts2014fig7}
\end{figure*}

The critical radius of meteoric smoke particles is different for neutral and charged meteoric smoke particles. \citet{Megneretal2008} modelled the critical radius for neutral and charged meteoric smoke particles as a function of temperature. They found that, in general, the critical radius is slightly lower for charged meteoric smoke particles. Importantly, the critical radius of charged particles vanishes for temperatures colder than \SI{126}{\kelvin}, assuming a water vapour mixing ratio of \SI{2}{\ppm} and a neutral air number density of $N=$\SI{1e20}{\per\cubic\metre}. That means, that below \SI{126}{\kelvin}, any charged meteoric smoke particle, regardless of its size, will act as an ice nucleus. For temperatures between \SIrange[range-phrase=\ and\ ]{126}{136}{\kelvin}, the critical radius increases with temperature from \SIrange[range-phrase=\ to\ ]{0.6}{2}{\nano\metre} \citet[][Fig. 5]{Megneretal2008}.

When the temperature is sufficiently low and ice nuclei are present, polar mesospheric clouds (noctilucent clouds) may form. Figure \ref{fig:baumgartenfritts2014fig7} shows photographs of these clouds made from the ground. \citet{BaumgartenFiedler2008} summarized the mean NLC particle radius and number density from measurements on And\o ya between 1998 and 2005. They found mean particle number densities of about \SIrange[range-phrase=\ to\ ]{30}{150}{\per\cubic\centi\metre}, and a mean radius\footnote{Note that the ice particles in general are not spherical \citep{Rappetal2007b}, and that the radius is often given as an effective radius rather than a true radius of a perfect sphere.} between approximately \SIrange[range-phrase=\ to\ ]{38}{59}{\nano\metre}. Generally, the radius increased towards the bottom of the NLC layer, while the particle number density decreased. This is expected, because the particles are believed to grow through coagulation while they sediment, thereby reducing the number of particles.

\citet{Megneretal2008} showed that meteoric smoke particle number densities in polar summer are far lower than observed NLC particle number densities. That is, the nucleation on neutral meteoric smoke particles cannot be considered the most likely case anymore. Their results confirm that charged meteoric smoke particles play a major role as condensation nuclei.

Unfortunately, as of today, no experiment has been conducted which has succeeded in sampling the ice particles and resolving the question of what the nucleus might be.

\subsection{Mesosphere to stratosphere}
The residual meridional circulation in the mesosphere transports these meteoric smoke particles to the winter pole, where they decend into the stratosphere and further down. \citet{Dhomseetal2013} estimated an atmospheric residence time of meteoric smoke particles of four to five years, i.e. from ablation until surface (or ocean) deposition. They assumed a mean radius of \SI{1.5}{\nano\metre} and a density of \SI{2000}{\kilogram\per\cubic\metre}.

As a result, the meteoric smoke particle concentration over the winter pole is larger than over the summer pole. Still, the meteoric smoke particles in the summer polar mesosphere are very important, because they are essential in allowing ice crystals to grow. 

\subsection{Stratosphere}

\citet[][and references therein]{Plane2003} proposed that meteoric smoke particles can affect the stratospheric ozone chemistry. As mentioned in Sect. \ref{sec:abundance}, the meteoric smoke particles are transported towards the winter polar mesosphere, where they sediment down into the winter stratosphere. The smoke particles have a large dipole moment, and can therefore react quickly with \ce{HCl}. Eventually, this reaction can lead to chlorine radicals that destroy ozone molecules. Meteoric smoke particles might also be important for the formation of nitric acid trihydrade $\left(\ce{HNO3.3H2O}\right)$, see \citet{Voigtetal2000}, for instance. Meteoric smoke particles may also remove sulphuric acid from the stratosphere above \SI{40}{\kilo\metre} \citep{Saundersetal2012}.

In the upper stratosphere (and possibly lower mesosphere), meteoric smoke particles are also likely condensation nuclei for \ce{H2SO4}, leading to the removal of neutral \ce{H2SO4} aerosols \citep{Hervigetal2017}. Analyzing SD--WACCM model runs, these authors showed that SOFIE observations of extinction can be explained better when these aerosols are included in the model \citep{Hervigetal2017}.

\subsection{Troposphere and Earth's surface}
After about four to five years since their formation, meteoric smoke particles are deposited at the Earth's surface \citep{Dhomseetal2013}. In their model study, \citet{Dhomseetal2013} showed that iron--rich particles can be deposited at midlatitudes, for example to the Southern Ocean, see Fig. \ref{fig:dhomseetal2013}. Compared to Aeolian dust, the meteoric smoke particles are relatively soluble, which compensates for their low mass flux to the surface compared to Aeolian dust \citep{Dhomseetal2013}. The dissolved iron can stimulate phytoplankton growth in the ocean.

\section{Detecting meteoric smoke particles}
\label{sec:detection}
Several measurement techniques have been used to measure the number density and charge of meteoric smoke particles, as well as to determine their composition. The techniques can be grouped into in situ measurements by sounding rockets and remote sensing by either ground-- or space--based instruments.

The rocket techniques have had two main objectives: particle detection and particle sampling. Remote sensing of meteoric smoke particles has mainly been done using satellite--based extinction measurements, which have attempted to deduce the composition of meteoric smoke particles, while ground--based incoherent scatter radar measurements (at Arecibo on Puerto Rico, and with the European Incoherent Scatter Radar) determined the altitude and size distributions of meteoric smoke particles (see Sect. \ref{sec:isr}.

I will focus on the results obtained with particle detection techniques with sounding rockets, and only briefly mention the results from satellite--based extinction measurements and from radar measurements.

\subsection{Rocket--borne techniques}

\subsubsection{Classical Faraday cup design}
\begin{figure*}[!t]
\centering
\includegraphics[width=0.7\textwidth]{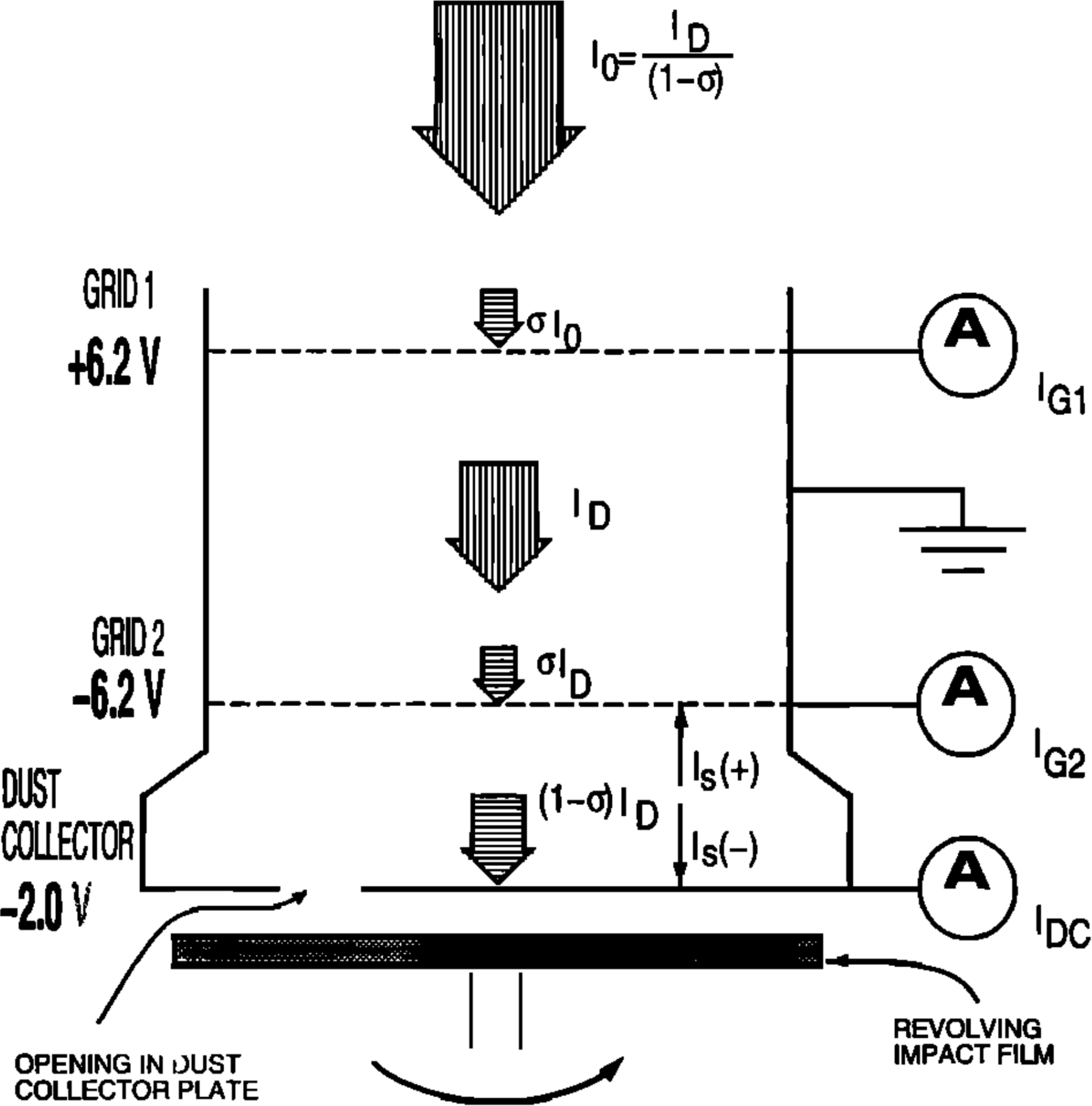}
\caption{Figure reprinted eproduced from \citet[][Fig. 1]{Havnesetal1996}.}
\label{fig:dusty}
\end{figure*}

In the 1990s, the DUSTY probe was developed and flown on rockets \citep{Havnesetal1996}. The design is sketched in Fig. \ref{fig:dusty}, and is often referred to as the ``classical Faraday cup design''. It may not come as a surprise that the DUSTY probe consists of a Faraday cup. The probe consists of a negatively biased grid, a positively biased grid, and the collector electrode. To improve aerodynamics (and collection efficiency, the Faraday cup has outflow holes behind the collector electrode.\\

The first, negatively biased, grid is held at a potential of \SI[explicit-sign=-]{6.2}{\volt}. This grid prevents free electrons and light negative ions from entering the Faraday cup. The second, positively biased, grid is held at \SI{6.2}{\volt} and stops light positive ions. The collector electrode is held at the payload potential.

As opposed to free electrons and ions, meteoric smoke particles are heavy enough to have a considerable kinetic energy. They can thus pass the two grids and reach the collector electrode. An impact on the collector electrode results in a measurable current.

The classical Faraday cup is a well--suited to measure the charge balance in combination with an electron probe and a positive ion probe. The downside is that one neither knows the charge state of a single particle nor the composition of the particles. One can only measure charged, not neutral, particles, and only those with an assumed radius of $r_{eff} > \SI{2}{\nano\metre}$.

If the payload then first ascends through the mesopause region and descends again, the instrument can detect whether there are charges present or not. By design, only one type of charge can be detected: positive or negative, dependent on the grid design. If a charge penetrates the upper grid and impacts on the detector, a current is measured.

A crucial assumption has to be made: each particle carries only a single charge. Furthermore, only particles larger than a certain radius can be detected \citep[][Fig. 2]{Rappetal2005}. This is because smaller particle are deflected away from the payload by aerodynamics. Using simulations, the effiency to detect meteoric smoke particles with radii between \SIrange[range-phrase=\ and\ ]{0.8}{1.4}{\nano\metre} dropped below \SI{50}{\percent} for common rocket velocities \citep{Horanyietal1999}. A similar results was shown by \citet[][Fig. 5]{Asmusetal2017}.This highlights how challenging even the mere detection of small particles is, let alone the sampling of those particles.

The classical Faraday cup has been a standard instrument on sounding rockets since its development.

\subsubsection{Detecting charged nanoparticles}
\begin{figure*}[!t]
   \centering
	 \includegraphics[width=0.5\textwidth]{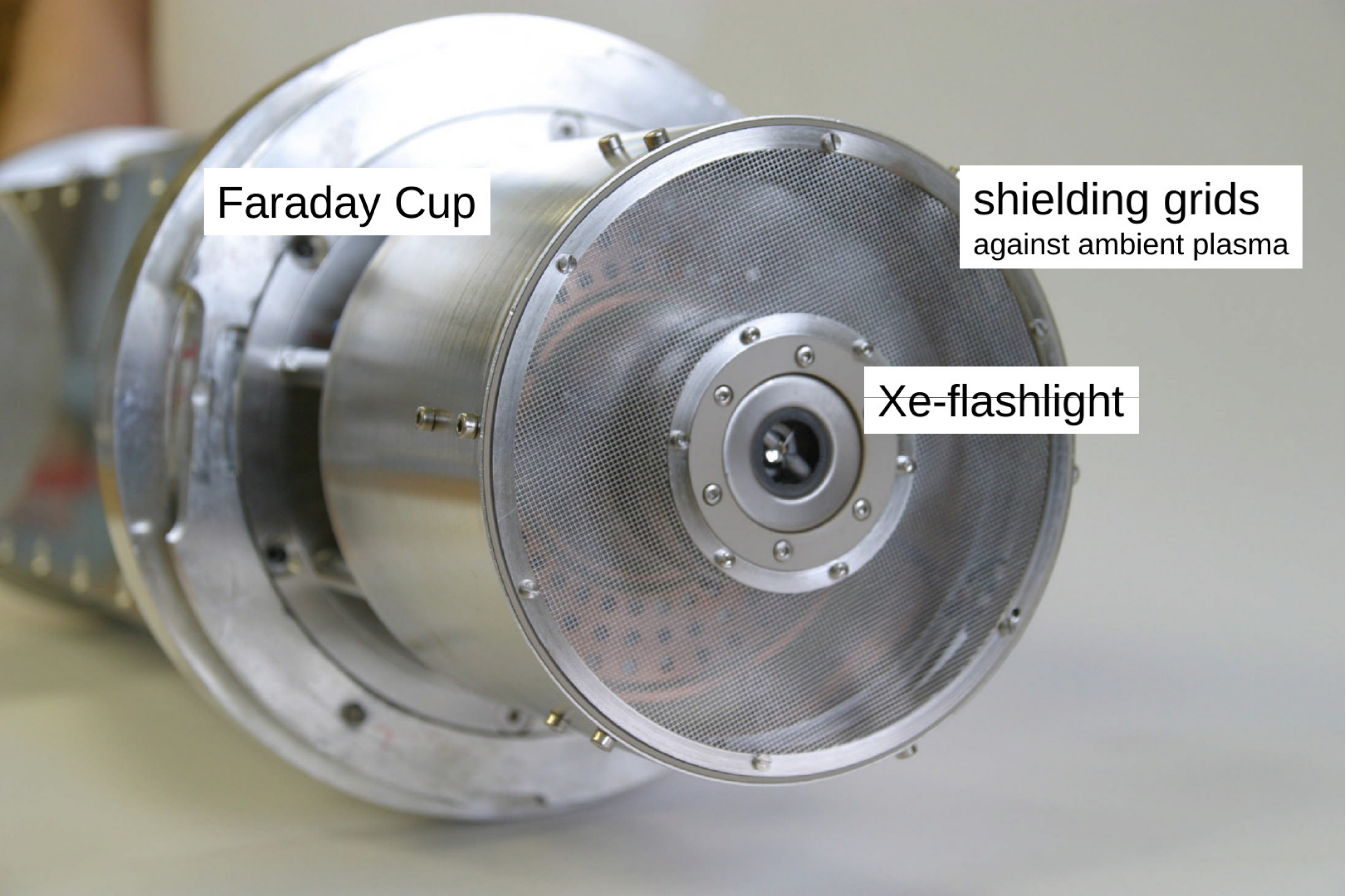}
	 \caption{Photograph of the modified ECOMA particle detector with a xenon ultraviolet flash lamp. From Rapp (pers. comm., 2013).}
	 \label{fig:ecoma}
\end{figure*}

\begin{figure*}[!t]
   \centering
	 \includegraphics[width=0.5\textwidth]{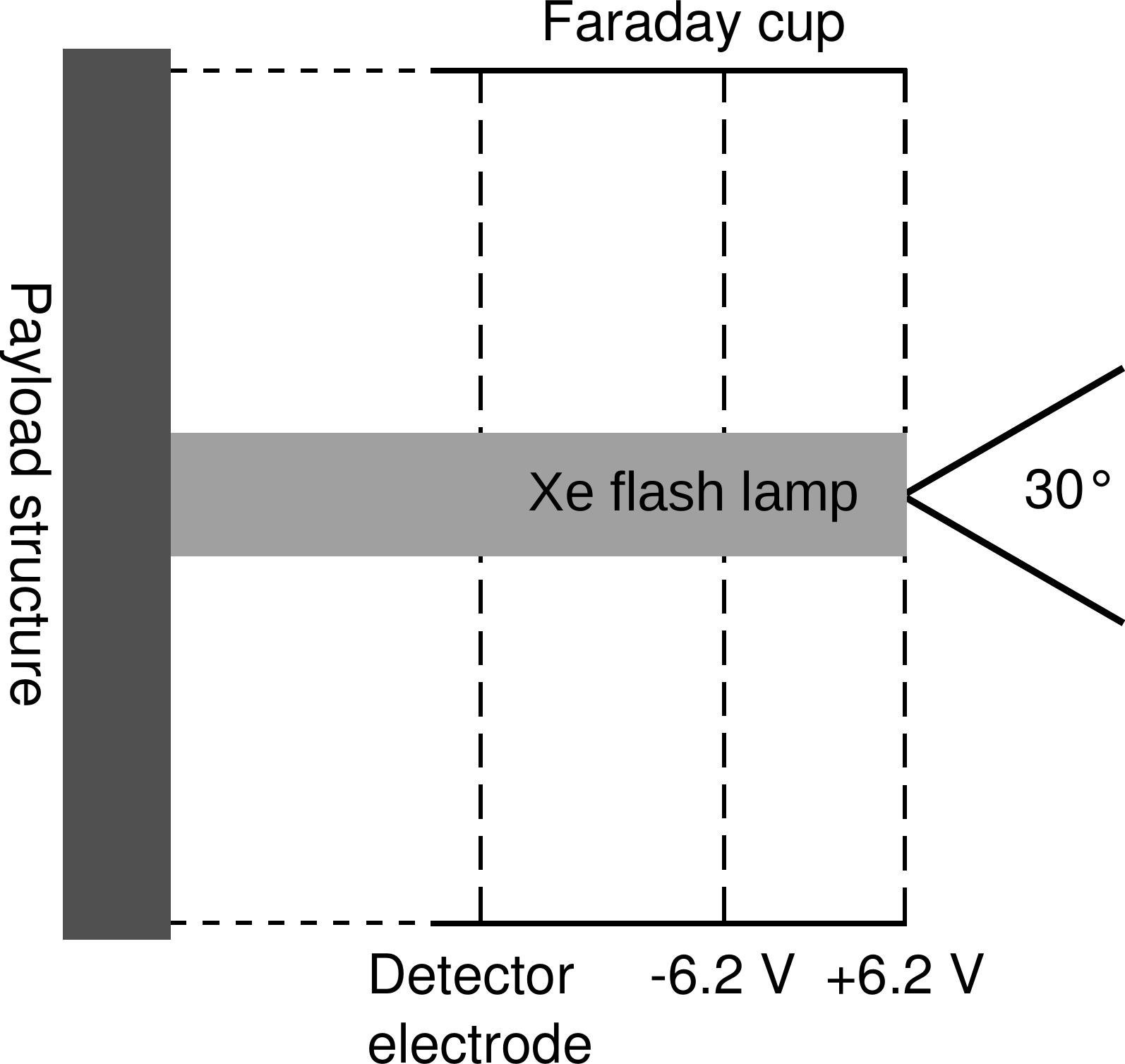}
	 \caption{Schematic view of the modified ECOMA particle detector with a xenon ultraviolet flash lamp, which emits pulses at a rate of \SI{20}{\hertz} at a full angle of \SI{30}{\degree}. A volume between \SIrange[range-phrase=\ and\ ]{2.5}{75}{\centi\metre} upstream of the payload is illuminated and particles  within this volume can interact with the photons emitted by the lamp. Adapted from \citet[][Fig. 1(b)]{RappStrelnikova2009}.}
	 \label{fig:ecoma_pd}
\end{figure*}

\citet{Rappetal2005} launched a rocket from Esrange in October 2004 in darkness. Aerodynamics prohibited the collection of particles smaller than $r_{eff}=\SI{2}{\nano\metre}$. They detected free electrons and positive ions, and in addition had a particle detector that detected positive particles which were not ions. The design is similar to that of \citet{Havnesetal1996}, who also describe the setup very well. A photograph and a sketch of the ECOMA particle detector is shown in Figs. \ref{fig:ecoma_pd} and \ref{fig:ecoma}.

\begin{figure*}[!t]
   \centering
	 \includegraphics[width=0.6\textwidth]{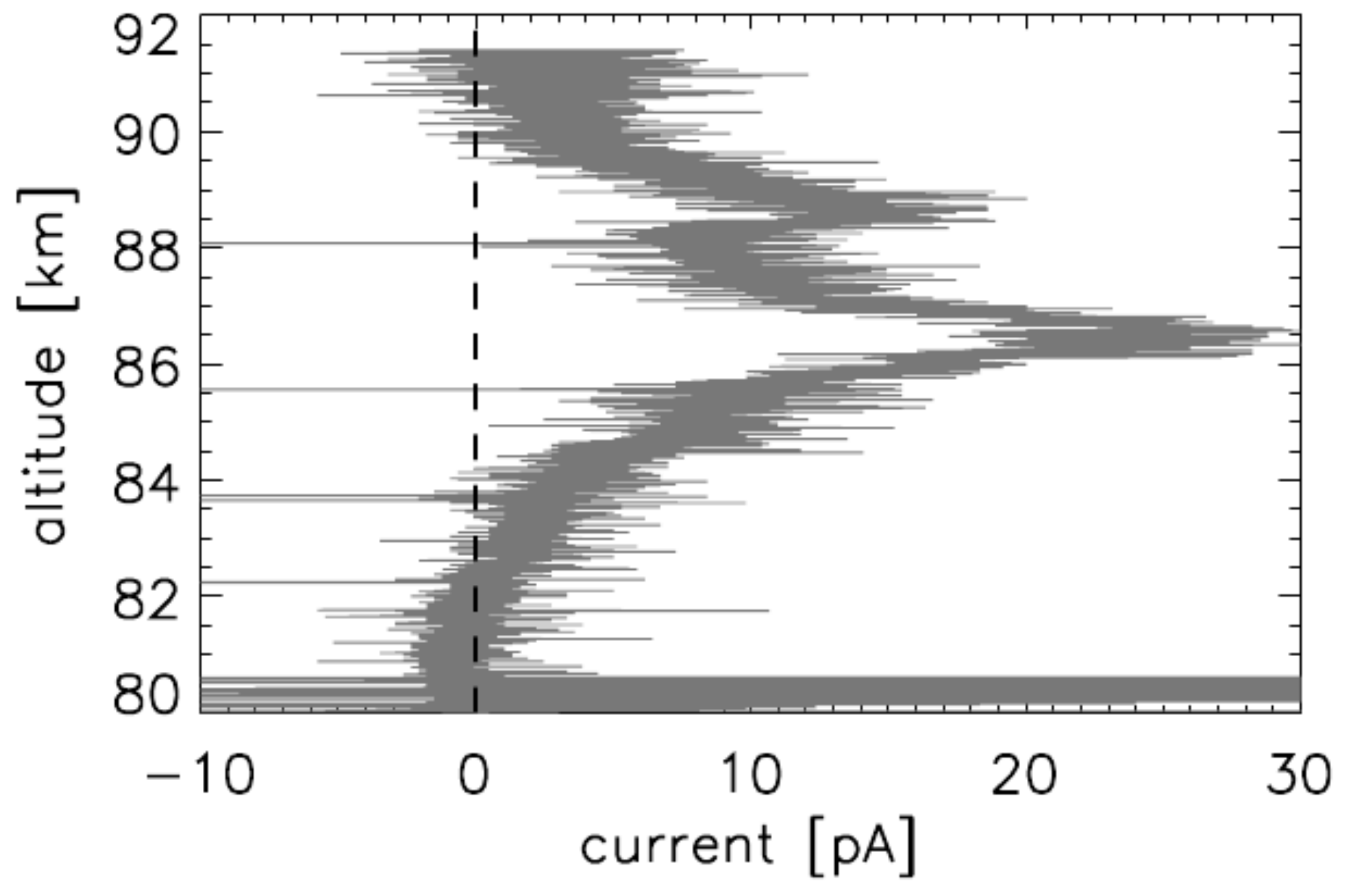}
	 \caption{Current measured during the rocket launch from Esrange in October 2004. Figure reprinted from \citet[][Fig. 4]{Rappetal2005}.}
	 \label{fig:rappetal2005fig4}
\end{figure*}

The measurement results are shown in Fig. \ref{fig:rappetal2005fig4}. The measured current can be converted into a charge number density by

\begin{equation}
   I = NZqAv_p,
\label{eq:rappetal2005}
\end{equation}
where $I$ is the measured current, $A$ is the particle detector area, $v_p$ is the payload's velocity relative to the particles, $N$ is the number density of particles, $q$ is the particle's charge, and $Z$ is the number of elementary charges per particle. Transformation of Eq. \eqref{eq:rappetal2005} yields
\begin{equation}
   NZq = \dfrac{I}{A v_p}
\end{equation}
Thus, one obtains a charge number density, $NZq$, without having to assume a certain number of elementary charges per particle. The maximum current was $I=\SI{25}{\pico\ampere}$, and the rocket velocity was $v_p \approx \SI{500}{\metre\per\second}$. The measurements yielded a maximum charge number density of about \SI{100}{\per\cubic\centi\metre} around \SI{86}{\kilo\metre}, which was produced by particles with radii equal to or larger than \SI{2}{\nano\metre} \citep{Rappetal2005}.

\subsubsection{Modified ECOMA particle detector}
\citet[][Sect.3.2]{Hedinetal2005} described the original ECOMA particle detector in some detail. The ECOMA particle detector is a modified version of the classical Havnes Faraday cup design described above. The modified detector includes three xenon ultraviolet flash lamps which emit ultraviolet light at different wavelengths.

From laboratory measurements, properties of some candidates for meteoric smoke particles are known. An important property is the work function, that is, the energy necessary to remove an electron from the particle. This energy is specific to the material. At a typical mesospheric temperature of $T=\SI{200}{\kelvin}$, the available thermal energy is
\begin{equation}
   \dfrac{3}{2}k_B T = \SI{4.14e-21}{\joule} = \SI{0.026}{\electronvolt}
\end{equation}
$\Rightarrow$ Much too small to remove an electron at typical mesospheric temperatures.

Therefore, another source of energy is necessary to remove electrons. One possible solution is a flashlamp that emits short high--energy pulses of light. The ECOMA particle detector uses an ultraviolet flashlamp. The UV flash lamps emit a rather broad spectrum, but with a sharp cut-off wavelengths that is determined by the window material. A volume between \SIrange[range-phrase=\ and\ ]{2.5}{70}{\centi\metre} upstream of the payload gets ionized by the flashes. The flash lamps have an opening angle of \SI{30}{\degree}, and operate at \SI{20}{\hertz} with a pulse energy of \SI{0.5}{\joule}. The maximum pulse energy (i.e., at the shortest possible wavelength) is thus $\sim$\SI{11.3}{\electronvolt}.

In principle, Eq. \eqref{eq:rappetal2005} also applies to the current measurements made with the xenon flash lamps, but the equation needs to be adapted to the instrument. For example, the flash lamp emits a spectrum instead of one discrete frequency. Therefore, one needs to integrate over the spectrum: from the cut-off wavelength to the work function wavelength. Furthermore, the flash lamp illuminates a certain volume, such that one also need to integrate over this volume. One also need to integrate over the assumed particle size. For more details, see \citet[][Eqs. (1) and (2)]{Rappetal2010}. The ECOMA particle detector can measure the volume number density. The number of electrons that can be attached to a particle should scale with its volume.

\subsubsection{Mass spectrometry}
\label{sec:massspec}
\begin{figure*}[!t]
  \centering
  \includegraphics[width=0.3\textwidth]{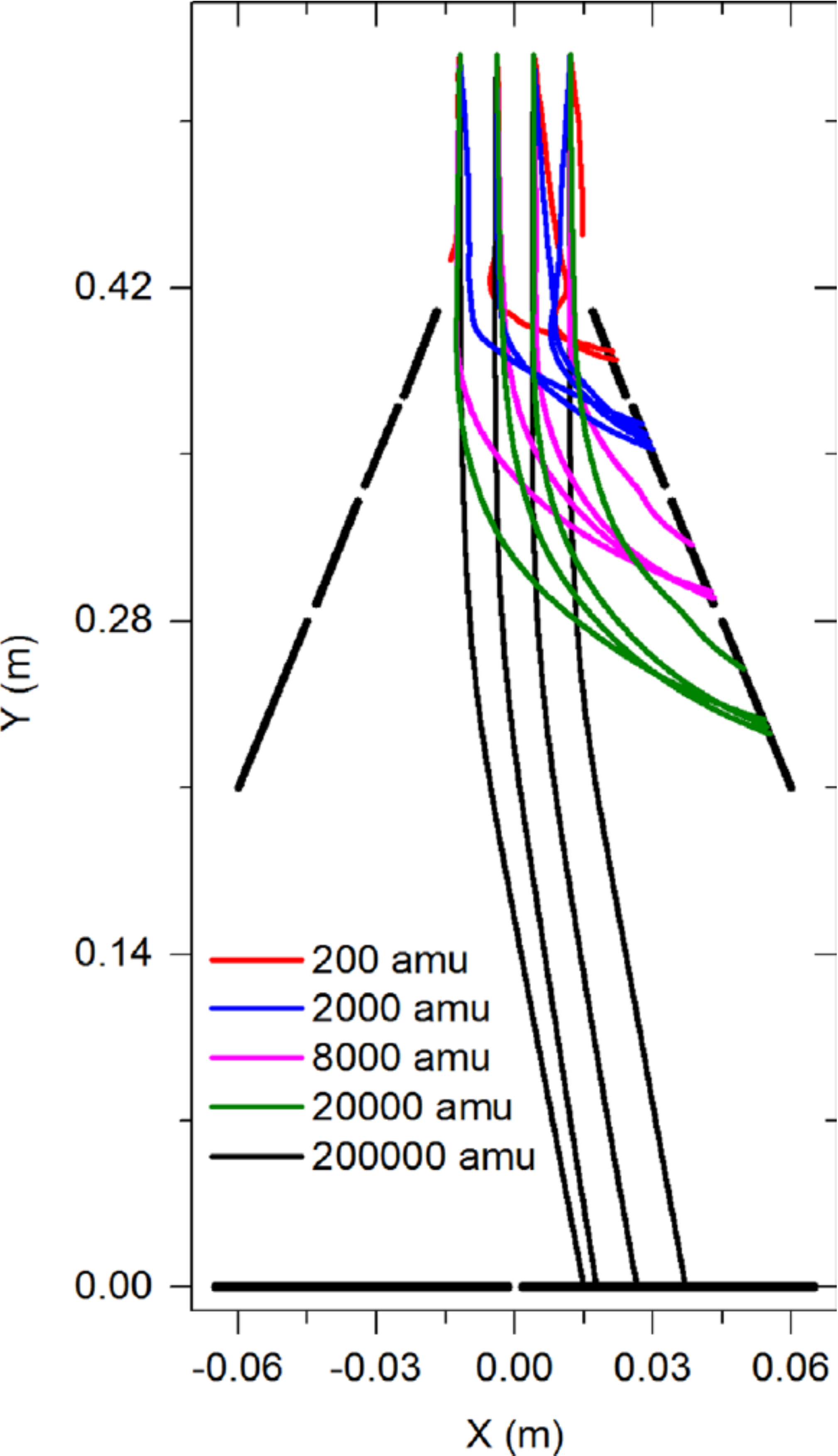}
  \caption{Sketch of the Mesospheric Aerosol Sampling Spectrometer. Depending on the particles' mass, their paths vary inside the detector, with the heaviest particles travelling furthest. Figure reprinted from \citet[][Fig. 4]{Robertsonetal2014}.}
  \label{fig:robertsonetal2014_1}
\end{figure*}

Mass spectrometers are, unsurprisingly, capable of resolving the mass of different particles. One such mass spectrometer that has been flown on sounding rockets is the Mesospheric Aerosol Sampling Spectrometer \citep[MASS;][]{Robertsonetal2014}, shown in Fig \ref{fig:robertsonetal2014_1}. It was mounted on the forward side of the rocket payload. On each side of the spectrometer, there are four biased detectors. One side has positively biased detectors, the other side has negatively biased detectors. This enables the mass--resolved measurement of both positive and negative charges. The opening area is about $A=\SI{25e-4}{\square\metre}$. The rocket velocity was $v_p=\SI{1050}{\metre\per\second}$. The measured current can again be converted to a number density using Eq. \eqref{eq:rappetal2005}, when one assumes that each particle carries only one charge:

\begin{equation}
   n=\dfrac{I}{Zqv_pA}
\end{equation}

The resolution of the instrument is \SI{1}{\pico\ampere}, with a noise level of \SI{8}{\pico\ampere}. This noise level corresponds to an uncertainty of the number density of $\sigma_n=\SI{20e6}{\per\cubic\metre}$. The charge is the elementary charge $e\approx\SI{1.602e-19}{\coulomb}$.

\begin{figure*}[!t]
  \centering
  \includegraphics[width=0.95\textwidth]{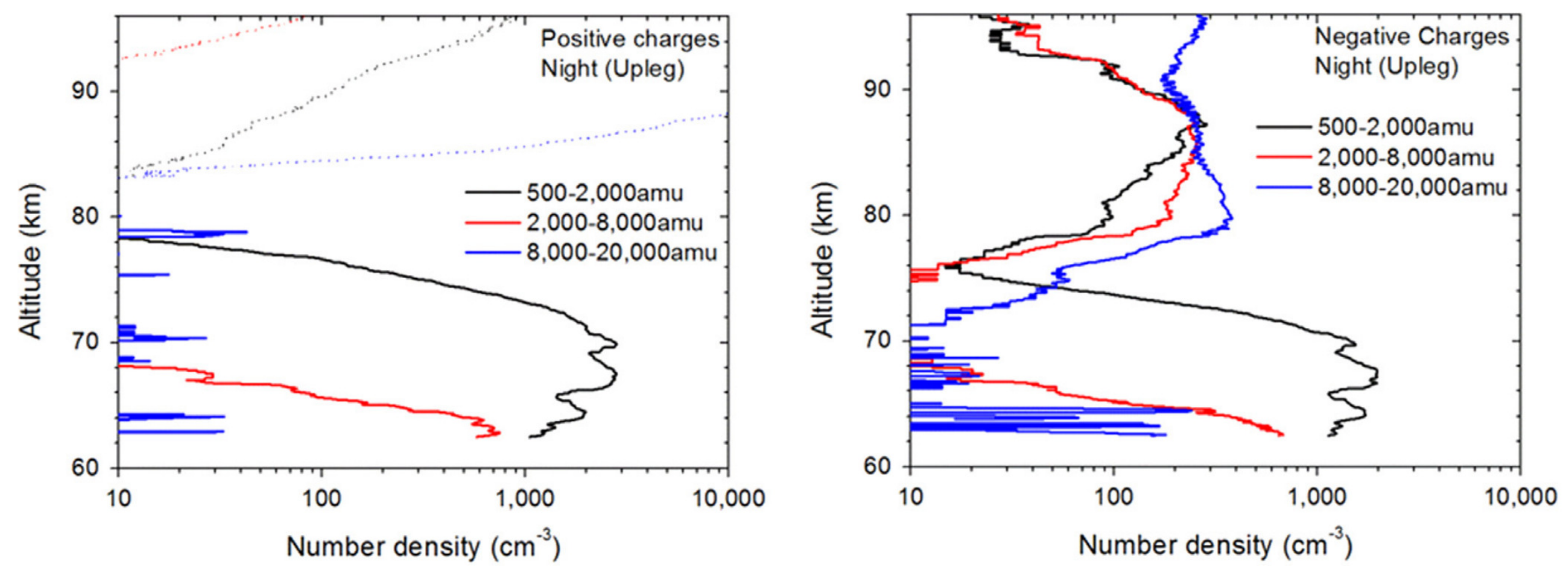}
  \caption{Left panel: altitude profile of meteoric smoke particle radius. Right panel: altitude profile of meteoric smoke particle number density. Spurious signal are shown as dotted lines. Figure reprinted from \citet[][Fig. 16]{Robertsonetal2014}.}
  \label{fig:robertsonetal2014_2}
\end{figure*}

Results from the Mesospheric Aerosol Sampling Spectrometer are shown in Fig. \ref{fig:robertsonetal2014_2}. Only results from measurements at night are shown here. Between \SIrange[range-phrase=\ and\ ]{78}{95}{\kilo\metre}, there were basically no positively charged meteoric smoke particles, only negatively charged particles, with the heaviest particles being most abundant. Below \SI{78}{\kilo\metre}, negatively and positively charged meteoric smoke particles coexisted with roughly equal number densities. The reason is that the electron density is greatly reduced below \SI{78}{\kilo\metre}. The results of \citet{Robertsonetal2014} confirm what \citet{Rappetal2005} already showed, namely the existence of positively charged nanoparticles in the nighttime polar mesosphere.

\subsubsection{Particle sampling}
Sampling (that is, collecting) neutral meteoric smoke particle is very challenging, as \citet{Hedinetal2014} pointed out in their summary of several attempts with the Mesospheric Aerosol -- Genesis, Interaction and Composition (MAGIC) particle detector. Whether particles can be collected, depends to a large degree on the sticking efficiency of the particles on the detector. Several different detector designs have been developed, always trying to increase this sticking efficiency.

The MAGIC detector is, in principle, capable of sampling neutral meteoric smoke particles, but the experiments turned out to be very difficult to perform. Interpretation of the data requires analysis of many instrumental parts, including the environment where the detector was manufactured and calibrated. \citet{Hedinetal2014} have not obtained conclusive results.

A new attempt to sample meteoric smoke particles is currently underway by Havnes and co--workers, with rocket launches from the And\o ya Space Center in northern Norway \citep[see][]{Havnesetal2015}.

\subsection{Satellite--based extinction measurements}
\label{sec:satellite}
The only satellite--based method that worked hitherto is the extinction measurement made by the Solar Occultation For Ice Experiment \citep[SOFIE;][]{Gordleyetal2009}. SOFIE, on board the AIM satellite, is in a polar orbit and measures extinction solely in the southern hemisphere \citep[e.g.,][]{Hervigetal2009a,Hervigetal2009b}.

\begin{figure*}[!t]
  \centering
  \includegraphics[width=0.48\textwidth]{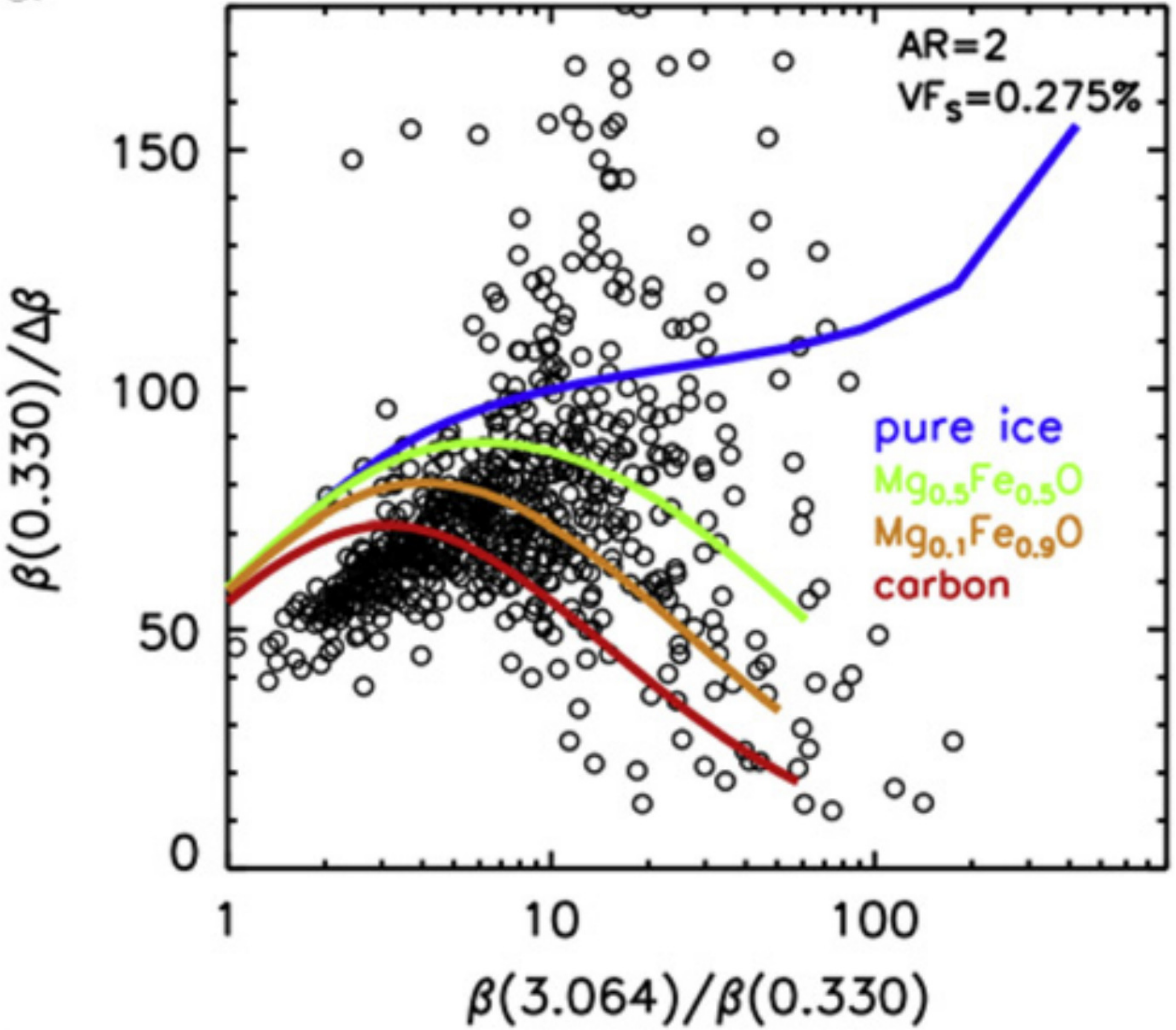}
  \includegraphics[width=0.48\textwidth]{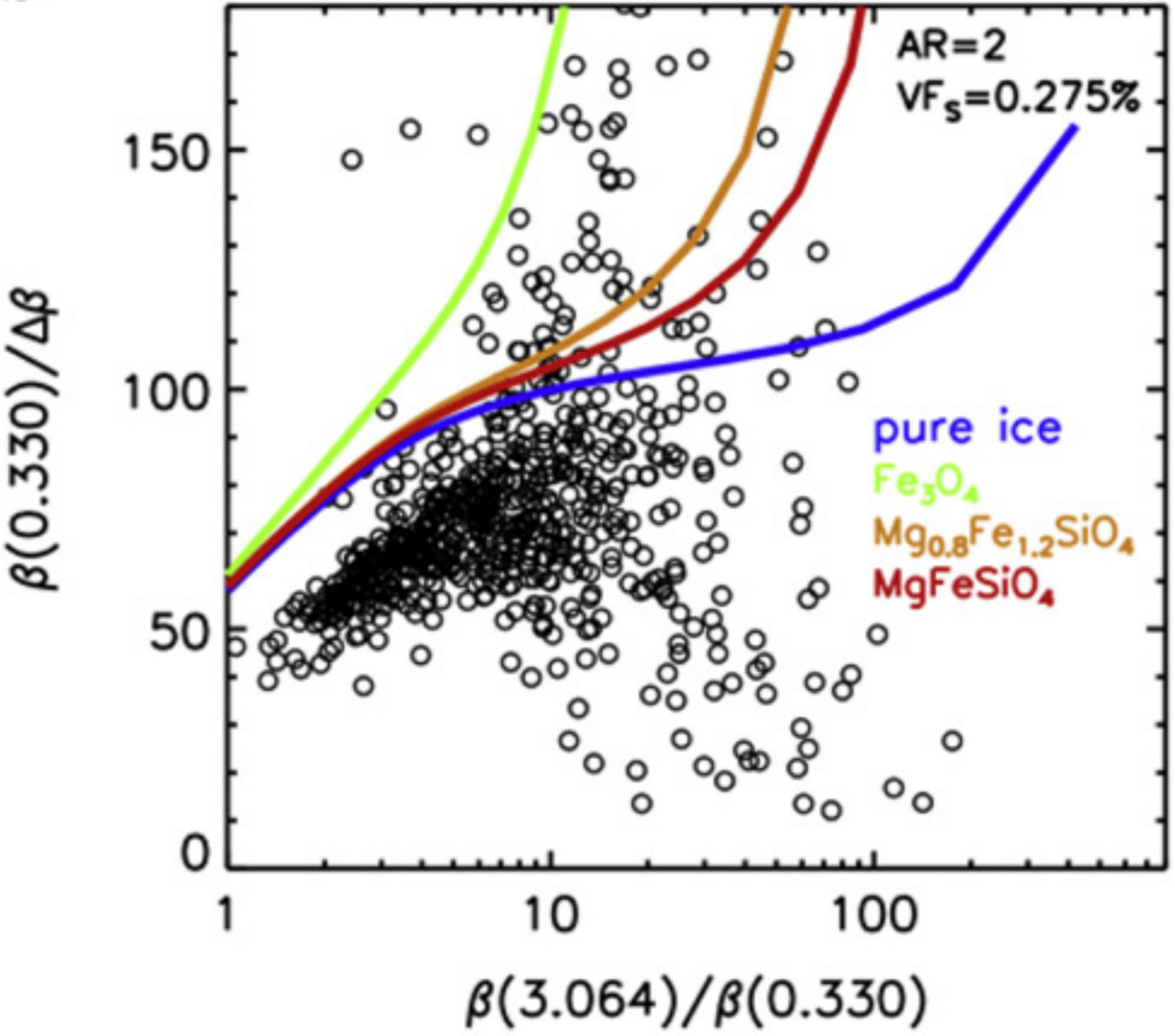}
  \caption{Extinction measurements made at different wavelengths by the Solar Occulation For Ice Experiment (black circles), and different modelled meteoric smoke particle properties fitted to the data (coloured circles). For the modelled data, a volume fraction of meteoric smoke in ice particles of $VF_S=\SI{0.275}{\percent}$ has been assumed, as well as an axial ratio of the ice particles of $AR=\num{2}$. (a) Probable composition of meteoric smoke particles that match the experimental data. (b) Composition of meteoric smoke particles that are not consistent with the observations. Note that pure ice or silicon content only explain a minor part of the observations. Figure reprinted from \citet[][Fig. 3]{Hervigetal2012}.}
  \label{fig:sofie}
\end{figure*}

Typically, an axial ratio for the particles has to be assumed. The axial ratio is defined as a particles diamter divided by its length. A perfectly spherical particles thus has an axial ratio of $AR=\num{1}$. The satellite measurements from the Solar Occultation For Ice Experiment usually assume $AR=\num{2}$, corresponding to elliptical particles.

\subsection{Incoherent scatter radar measurements}
\label{sec:isr}
To derive the meteoric smoke particle size and number density, \citet{Strelnikovaetal2007} and \citet{Rappetal2007a} used an incoherent scatter radar. \citet{Choetal1998} laid the theoretical framework for incoherent scatter radar measurements\footnote{See also \citet{LaHoz1992}, who calculated how the radar spectrum would change in the presence of charged particles. His results showed that it is possible to detect such particles with radars, because there should be an enhancement in radar backscatter, its strength depending on the charge number and the frequency at which the radar operates \citep{LaHoz1992}}. They showed how the presence of charged meteoric smoke particles can change the shape of the Doppler spectra measured by such radars: in general, meteoric smoke particles in the atmosphere lead to a narrower Doppler spectrum compared to an atmosphere without such particles \citep{Choetal1998}. A special case is the presence of very small, negative meteoric smoke particles, which widen the Doppler spectrum \citep{Choetal1998}.

\begin{figure*}[!t]
  \centering
  \includegraphics[width=0.7\textwidth]{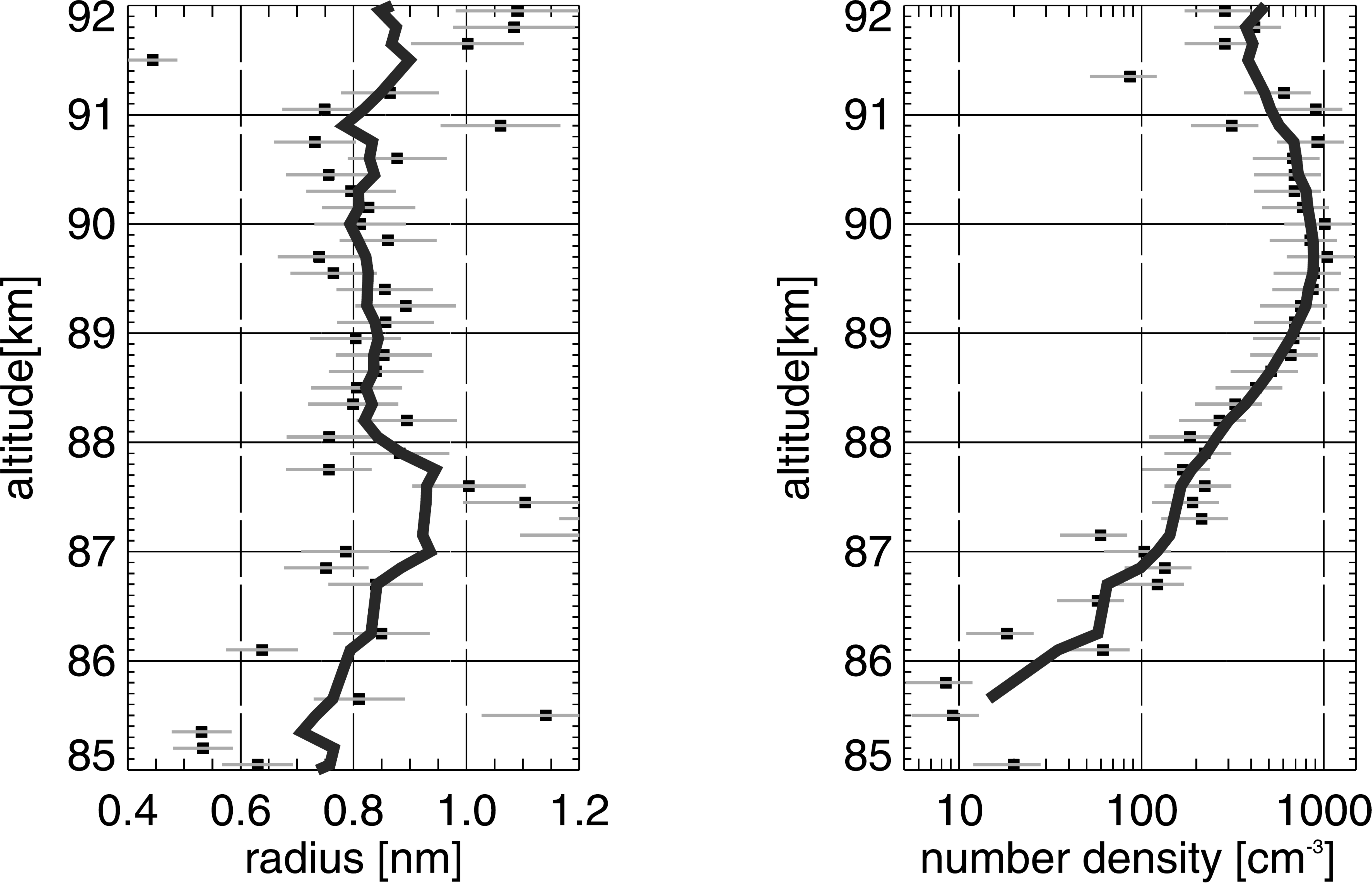}
  \caption{Meteoric smoke particle size and number density measurements made by the Arecibo incoherent scatter radar. Left panel: altitude profile of meteoric smoke particle radius. Right panel: altitude profile of meteoric smoke particle number density. Figure reprinted from \citet[][Fig. 4]{Strelnikovaetal2007}.}
  \label{fig:arecibo}
\end{figure*}

Figure \ref{fig:arecibo} shows one example of meteoric smoke particle number density and size, measured by \citet{Strelnikovaetal2007} with the Arecibo incoherent scatter radar on Puerto Rico. The radar operated at \SI{430}{\mega\hertz}.

To the best of my knowledge, no incoherent scatter radar measurements of meteoric smoke particles newer than the cited ones have been published, despite the success of this technique. Incoherent scatter radar measurements can be done much more regularly than sounding rocket flights, and are also independent of weather conditions. Suitable incoherent scatter radars exist at several locations. Thus, the profiles of number density and particle size can be measured throughout the year, with the potential to reveal seasonal and latitudinal differences. 

\section{What can we learn from it all?}
Meteoric smoke particles are likely the ablation products with the largest effect on the neutral and charged atmosphere. I have neglected meteoroids so small that they never reach their sublimation temperature, and those meteoroids are so large that they pass the middle atmosphere almost unaffected.

Further, I have looked in some detail at sounding rocket techniques to detect several properties of meteoric smoke particles. In brief, I have also described how satellite remote sensing can be used to infer particle properties. I have not covered spectroscopy methods and radar remote sensing. For information on radar remote sensing of meteoric smoke particles, see \citet{Rappetal2007a} and \citet{Strelnikovaetal2007}.

There is today not one single technique that can do it all, and there will probably never be such a technique. The most useful ``technique'' is actually a combination of as many different instruments as possible, because every instrument has its advantages and disadvantages. 

The composition of these particles is still unresolved, even though both rocket and satellite data have pointed to a composition mainly consisting of metal oxides and metal hydroxides. Silicon content seems to only play a minor role. The charge state of meteoric smoke particles is extremely important for the charge balance of the lower ionosphere and for the nucleation of ice particles. Neutral particles are probably too few to explain the occurrence of a much larger number of ice particles. Charged smoke particles decrease the energy barrier for nucleation. Even more important, below a certain temperature, any charged meteoric smoke particle can act as an ice nucleus, independent of its size.

With the exception of satellite--based measurements, observations are rather scarce and thus were conducted under different geophysical conditions, e.g. night and day, summer and winter, and latitude. The possibly different geophysical conditions have to be considered when different experimental results are compared to each other.

\section{Current research questions}
Finally, some topical research questions are:
\begin{enumerate}
 \item What is the composition of meteoric smoke particles? How important is silicon?
 \item How are meteoric smoke particles charged?
 \item Which role do meteoric smoke particles play for the occurrence of polar mesospheric winter echoes?
 \item How do meteoric smoke particles affect ozone chemistry in the stratosphere?
\end{enumerate}

\bibliography{bibliography_trial_lecture.bib}

\end{document}